\RequirePackage{fix-cm}
\documentclass[smallextended]{svjour3}       
\smartqed  
\usepackage{graphicx}
\usepackage{color}
\begin{document}

\title{Geodesic Motion in Schwarzschild Spacetime Surrounded by Quintessence
}

\author{Rashmi Uniyal         \and
N. Chandrachani Devi \and
 Hemwati Nandan \and
 K. D. Purohit 
}


\institute{   Rashmi Uniyal \at
              Department of Physics, Gurukul Kangri Vishwavidyalaya, Haridwar 249 407, India\\
              \email{rashmiuniyal001@gmail.com}           
           \and
              N. Chandrachani Devi \at
              Observat'orio Nacional, 20921-400, Rio de Janeiro - RJ, Brasil\\
              \email{chandrachani@on.br}
           \and
              Hemwati Nandan \at
              Department of Physics, Gurukul Kangri Vishwavidyalaya, Haridwar 249 407, India\\
              \email{hnandan@iucaa.ernet.in}
           \and
               K. D. Purohit \at
               Department of Physics, HNB Garhwal University, Srinagar Garhwal 246 174, India\\
               \email{kdpurohit@rediffmail.com} 
}

\date{Received: date / Accepted: date}

\maketitle

\begin{abstract}
We study the time-like geodesic congruences, in
the space-time geometry of a Schwarzschild black hole surrounded by quintessence.
The nature of effective potential along with the structure of the possible orbits for test
particles in view of the different values of quintessence parameter are analysed in detail.
An increase in quintessence parameter is seen to set the particles from \textbf{further} distance into motion around black hole.
The effect of quintessence parameter is investigated analytically wherever possible otherwise we perform the numerical analysis to probe the structure of possible orbits.
It is observed that there exist a number of different possible orbits for a test particle in case of non-radial geodesics, such as circular (stable as well as unstable) bound orbits, radially plunge and fly-by orbits, whereas no bound orbits exist in case of radial geodesics.

\textit{pacs} {04.70.Bw, 97.60.Lf, 95.36.+x}
\end{abstract}

\section{\label{sec:level1}Introduction}
The current cosmological observation from supernovae type Ia (SNe Ia), the cosmic microwave background (CMB), baryon acoustic oscillations (BAO) and Hubble measurements predicts that our universe is going through a phase of accelerating expansion, favoring to the existence of some unknown form of energy with a large negative pressure, named as dark energy. To determine the nature behind this unknown energy and to analyse its consequences on the other observable quantities, has become one of the most fascinating tasks at present among the groups working mainly in cosmology and particle physics.
There are several candidates for dark energy, such as cosmological constant \cite{Padmanabhan:2002ji,Cunha:2003vg}, phantom \cite{Caldwell:1999ew,Chimento:2003qy,Caldwell:1997ii,Sahni:1999qe}, quintessence \cite{Capozziello:2005ra,Vikman:2004dc,rog}, K-essence \cite{Chiba:1999ka,Scherrer:2004au}
and quintom \cite{Wei:2005nw,Zhao:2005vj,f,Chimento:2008ws}.
 Basically, the difference between these candidates for dark energy lies in the magnitude of equation of state parameter ($\epsilon$) which is the ratio of pressure to energy
density of dark energy.
 In recent years, several models  attributed to the presence of mysterious dark energy \cite{Copeland:2006wr,Babichev:2014lda} and Quintessence are proposed \cite{DeFelice:2010aj,Nojiri:2006su,Borowiec:2006qr,Durrer:2007re,Nojiri:2006ri,Capozziello:2003tk,Capozziello:2007ec,Sami:2007zz}.
The simplest and the most consistent one with the observations is the vacuum energy model (i.e. $\epsilon = -1$),
which is also termed as the cosmological constant ($\Lambda$) model.
 However, due to some of the theoretical issues namely the fine-tuning and cosmic coincidence problems associated to this model, a wide range of alternative scenarios has been proposed time and again in the literature.
  Among such alternative models, the quintessence scalar field model, whose equation of state varies slowly with the cosmic expansion and mimics $\Lambda$, remains the most popular one with the $\epsilon$ lying in the range of $-1\leq\epsilon\leq-1/3$.
 It would therefore be interesting to investigate the motion of test particles in the background of a Schwarzschild black hole spacetime surrounded by the quintessence.
  Although, the effect of this dark energy is much negligible at our local universe, but its existence cannot be denied to have an impact on our universe at any scale.\\
 In General Relativity (GR), the curvature and geometry of the space-time play crucial role as space-time is curved with the presence of
matter fields \cite{Misner,hartle,toolkit,wald}.
A number of studies related to the geodesic motion in the background of various spacetimes has been performed time and again due to its astrophysical importance \cite{e1,Hackmann:2008zz,Hackmann:2008zza,Jamil:2014rsa,Diemer:2014lba;Grunau:2013oca;Grunau:2010gd,Fernando:2012ue}.
 In general, the effects of the
curvature in a given space-time is studied through the Geodesic Deviation Equations (GDE) \cite{Synge:1934zza,Pirani:1956tn,elli}, the equations which describe
the relative acceleration of two neighbouring geodesics in diversified scenario \cite{toolkit,wald,Ghosh:2009ig,Koley:2003tp,Uniyal:2014oaa,vb}.
So, the GDE not only provide us an elegant description about the
structure of a space-time, but also all the important relations (mainly Raychaudhuri equation \cite{Raychaudhuri:1953yv,Kar:2006ms,Kar}, appropirate Mattig relation, etc)
can be obtained by solving the GDE for the time-like, null and space-like geodesic congruences.\\
 In the present work, we consider the evolving quintessence scalar field dark energy model and study the geodesics around a Schwarzschild black hole surrounded by such scalar field.
  In particular, the study of geodesic structures and nature of effective potential for radial and non-radial geodesics with different values of quintessence parameter are discussed in Section II in detail.
  The behaviour of orbits of a test particle is then analysed for different shapes of effective potential accordingly in Section III. Section IV  is mainly concerned about the geodesic deviation, in which the behaviour of geodesic deviation vector along a time-like geodesic near singularities is investigated and the nature of tidal effects between nearby test particles in the above mentioned space-time geometry are examined. We summarise our results in Section V.

\section{\label{sec:level1}The quintessence Schwarzschild black hole space-time}
We consider a Schwarzschild black hole surrounded by \textbf{quintessence} scalar field whose equation of state parameter is given by
\begin{equation}
 \epsilon = \frac{p_{\Phi}}{\rho_{\Phi}} = \frac{\frac{1}{2}\dot{\Phi}^2-V(\Phi)}{\frac{1}{2}\dot{\Phi}^2+V(\Phi)},
 \end{equation}
where the pressure $p_{\Phi}$ and  $\rho_{\Phi}$ are defined in terms of the kinetic term (i.e. $\dot{\Phi}^2$) and potential energy $V(\Phi)$  of scalar field as $\frac{1}{2}\dot{\Phi}^2-V(\Phi)$ and $\frac{1}{2}\dot{\Phi}^2+V(\Phi)$, respectively. Here, dot represents the differentiation with respect to the cosmic time. For a slowly rolling quintessence field,  $\dot{\Phi} \ll  V(\Phi)$, it exactly mimics to the cosmological constant model, $\epsilon \sim -1$.  For the static spherically-symmetric quintessence surrounding a black hole, as investigated in Kiselev\cite{Kiselev:2002dx}, the energy density of quintessence field reduces to a form:
\begin{equation}
\rho_{\Phi} = - \frac{\alpha}{2}\frac{3\epsilon}{r^{3(1+\epsilon)}},
\end{equation}
where the values of $\epsilon$ lie in the range of $-1<\epsilon<-\frac{1}{3}$.
Here $\alpha$ is the normalization factor. With the fact that the energy density of scalar field, $\rho_{\Phi}$ is always a positive quantity and $\epsilon$, a negative value, the normalization factor $\alpha$ has to be a positive value as we have already introduced a negative signature in the expression of $\rho_{\Phi}$. Further the quintessence field is assumed as a barotropic fluid with constant equation of state with the given range of permissible values for parameter $\epsilon$. Based on such standpoints, the metric of Schwarzschild black hole acquires the following form,
\begin{equation}
ds^2 = f(r)dt^2-\frac{1}{f(r)}dr^2 -r^2(d\theta^2 + {\sin}^2 \theta d\phi^2)
\label{lineelement}
\end{equation}
where $$f(r) = 1-\frac{2M}{r}-\frac{\alpha}{r^{3\epsilon+1}},$$
with the black hole mass, M. One recovers the Schwarzschild black hole in the limit of $\alpha =0$, whereas with $\epsilon = -1$ the metric reduces to the Schwarzschild black hole with cosmological constant.
The geodesic equations and its constraint equations are given by,
\begin{equation}
  {\ddot{x}}^{\mu}+{{\Gamma}^{\mu}_{\nu\lambda}}{\dot{x}}^{\nu}{\dot{x}}^{\lambda}=0\label{eq:geo_eq1},
\end{equation}
\begin{equation}
   g_{\mu\nu}{\dot{x}}^{\mu}{\dot{x}}^{\nu}=e.
   \label{eq:const_eq1}
\end{equation}
Here dot denotes the differentiation with respect to the affine parameter $\tau$ and $x^{\mu}$ being the space time coordinates.
One can set $e = 0$ or $1$, which corresponds to null or timelike geodesics respectively.
The geodesic equations for the metric we considered take the following forms
\begin{equation}
\ddot t  + \frac{ f'(r)} {f(r)}\, \dot r \, \dot t  = 0,
\label{tg11}
\end{equation}
\begin{equation}
\ddot r  + \left (\frac{ f'(r) \, \dot t^2 +  {f'(r)}^{-1} \,\dot r^2  - 2 r \, \dot \theta^2 -2 r \, \sin^2 \theta \, \dot \phi^2} {2 \, f^{-1} (r)} \right ) = 0,
\label{rg11}
\end{equation}
\begin{equation}
\ddot \theta + \frac{2}{r}\, \dot r \, \dot \theta - \cos \theta \, \sin \theta\,  \dot \phi^2  = 0,
\label{ps1}
\end{equation}
\begin{equation}
\ddot \phi + \frac{2}{r} \, \dot r \, \dot \phi + 2 \, \cot \theta \, \dot \theta \,  \dot \phi =0,
\label{ph1}
\end{equation}
where the prime denotes the differentiation with respect to $r$. The time-like constraint on the trajectories is given by

\begin{equation}
\left(1-\frac{2M}{r}-\frac{\alpha}{r^{3\epsilon+1}}\right)\dot t^2- 
\left(1-\frac{2M}{r}-\frac{\alpha}{r^{3\epsilon+1}}\right)^{-1}\dot r^2
\nonumber
\end{equation}
\begin{equation}
 \hspace{3cm}-r^2(\dot \theta^2 + {\sin}^2 \theta \dot \phi^2) = 1.
\label{eq:const_eq2}
\end{equation}

Using above mentioned eqs. (\ref {tg11})-(\ref{eq:const_eq2}), one can easily study the behaviour of geodesic equations on the equatorial plane like in case of Schwarzschild black hole.

\subsection{\label{sec:level2}Geodesic equations on the equatorial plane and effective potential}
We consider the equatorial plane (i.e. $\theta= \pi/2$) and with this, one can integrate eq. (\ref{tg11}) and eq. (\ref{ph1}) which leads to:
\begin{equation}
\dot{t} = \frac{C_1}{1-\frac{2M}{r}-\frac{\alpha}{r^{3\epsilon+1}}},
\label{eq:tg2}
\end{equation}
\begin{equation}
\dot{\phi} = \frac{C_2}{r^2},
\label{eq:ph2}
\end{equation}
where the integrating constants $C_1 $ and  $C_2$ correspond to the conserved total energy $E$ and the conserved angular momentum $L$ of a test particle respectively. Substituting the above eq. (\ref{eq:tg2}) and eq. (\ref{eq:ph2}) along with $\theta =\pi/2$ in the constraint eq. (\ref{eq:const_eq2}), the energy conservation equation for the time-like geodesic reads as
\begin{equation}
{E^2}={\left({\frac{dr}{d\tau}}\right)^2}+{V_{eff}}
\label{eqmotion}
\end{equation}
 where ${V_{eff}}$ is defined as an effective potential and is expressed as:
\begin{equation}
{V_{eff}} (r)= \left(1-\frac{2M}{r}-\frac{\alpha}{r^{3\epsilon+1}}\right)\left(\frac{L^2}{r^2}+1\right)\nonumber
\end{equation}
\begin{equation}
= \left(1-\frac{2M}{r}\right) +\frac{L^2}{r^2} -\frac{2ML^2}{r^3}-{\frac{\alpha}{r^{3\epsilon+1}}}\left(1+\frac{L^2}{r^2}\right).
\label{potential}
\end{equation}

\begin{figure}
\centering
\includegraphics*[width=5.6cm,height=4.5cm]{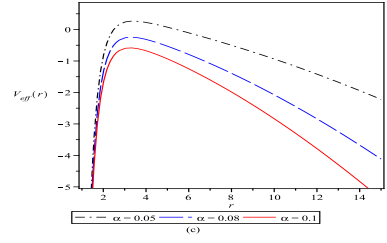}
\label{fig:p_4}
\caption{Behavior of the effective potential for different values of normalisation parameter $\alpha$, (label in the figure), with a unit black hole mass, $M=1$, a particular value of angular momentum (say ${L^2}=20$) and for different values of equation of state parameter,
 $(a)$: $\epsilon=-1/3$, $(b)$: $\epsilon=-2/3$
and
$(c)$: $\epsilon=-1$ respectively.}
\label{fig:pp_1}
\end{figure}

\begin{figure}[ht]
\centering
\includegraphics*[width=5.6cm,height=4.3cm]{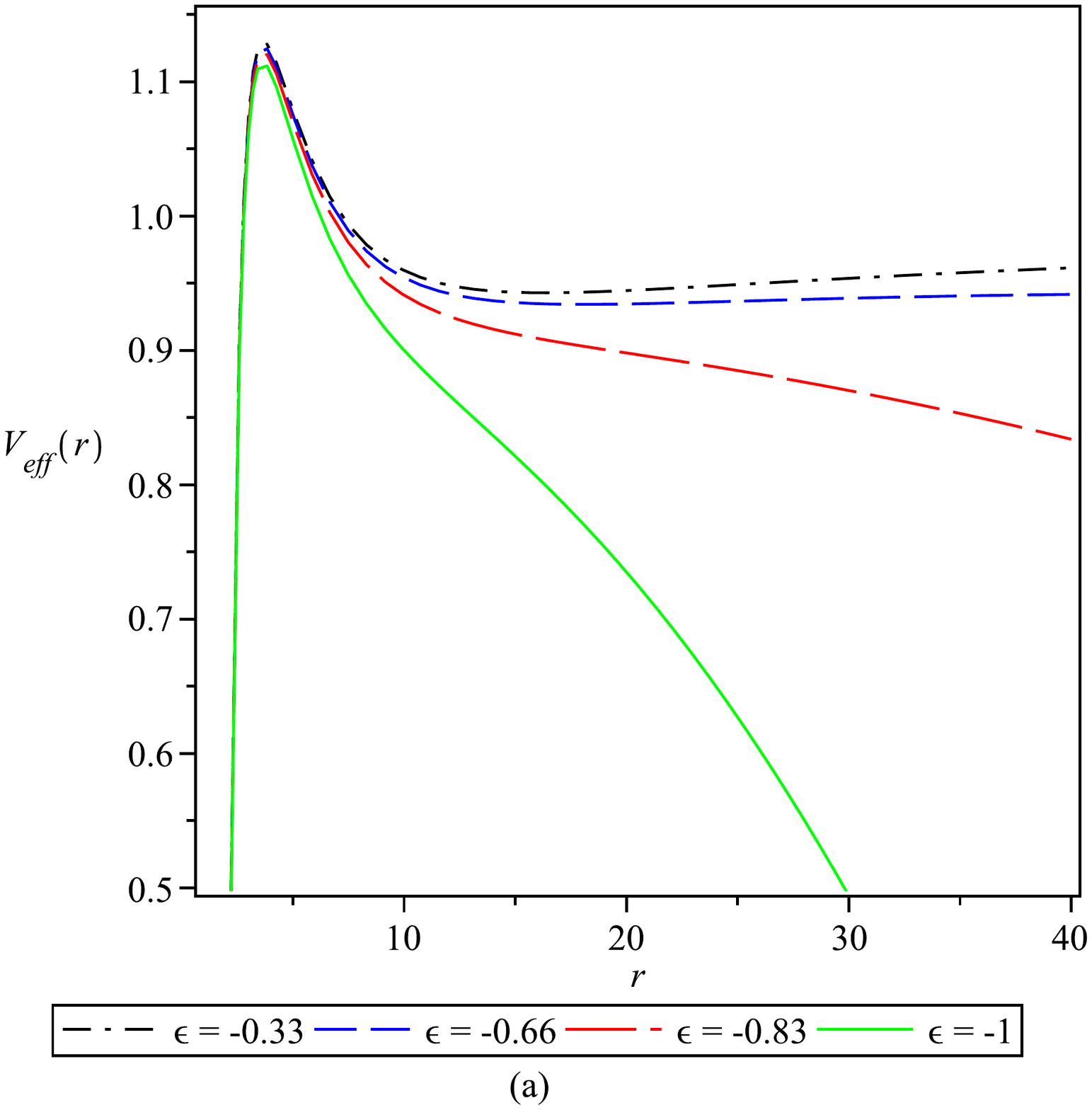}
\label{fig:p_6}
\includegraphics*[width=5.6cm,height=4.3cm]{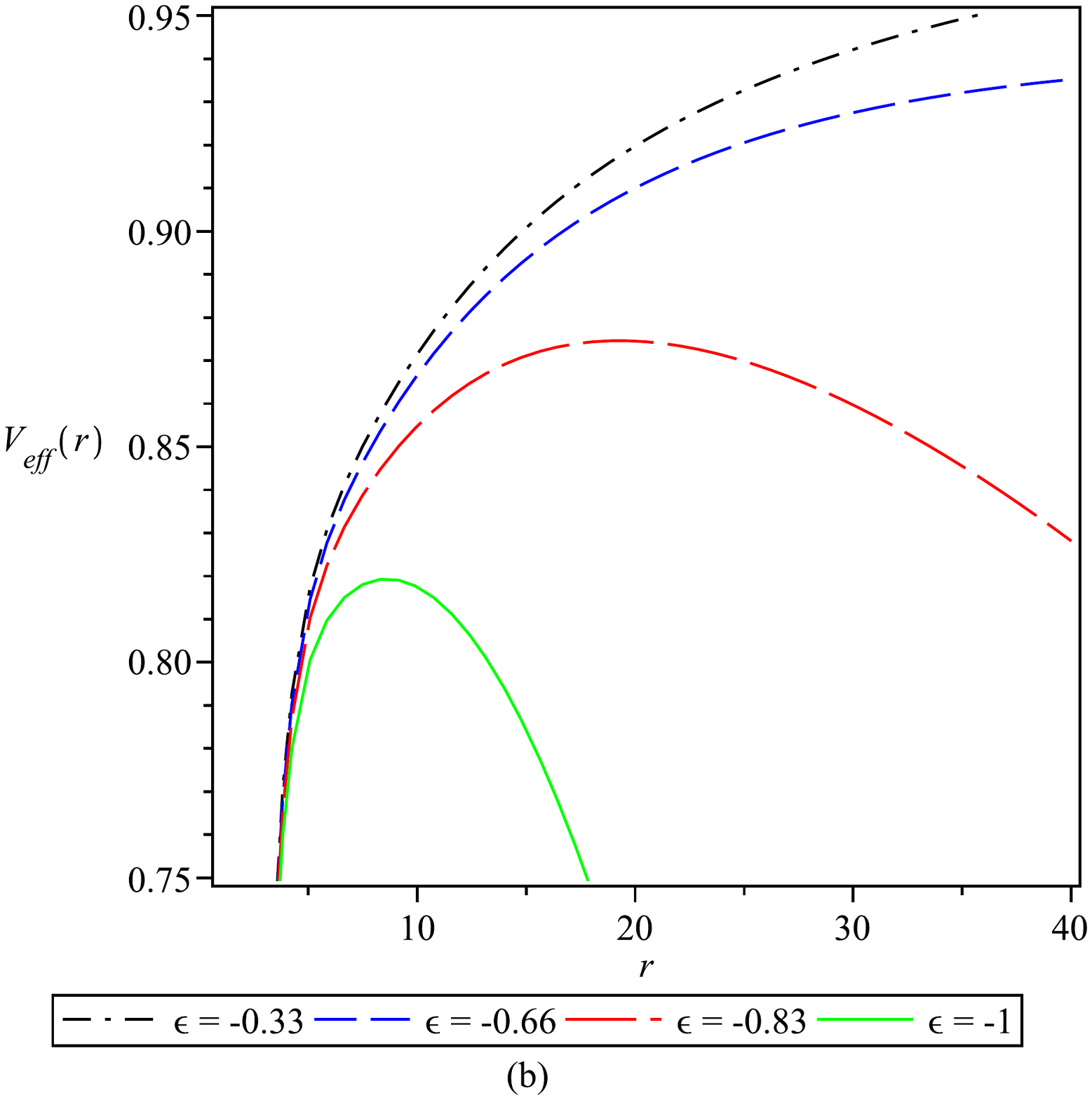}
\label{fig:p_7}
\includegraphics*[width=5.6cm,height=4.3cm]{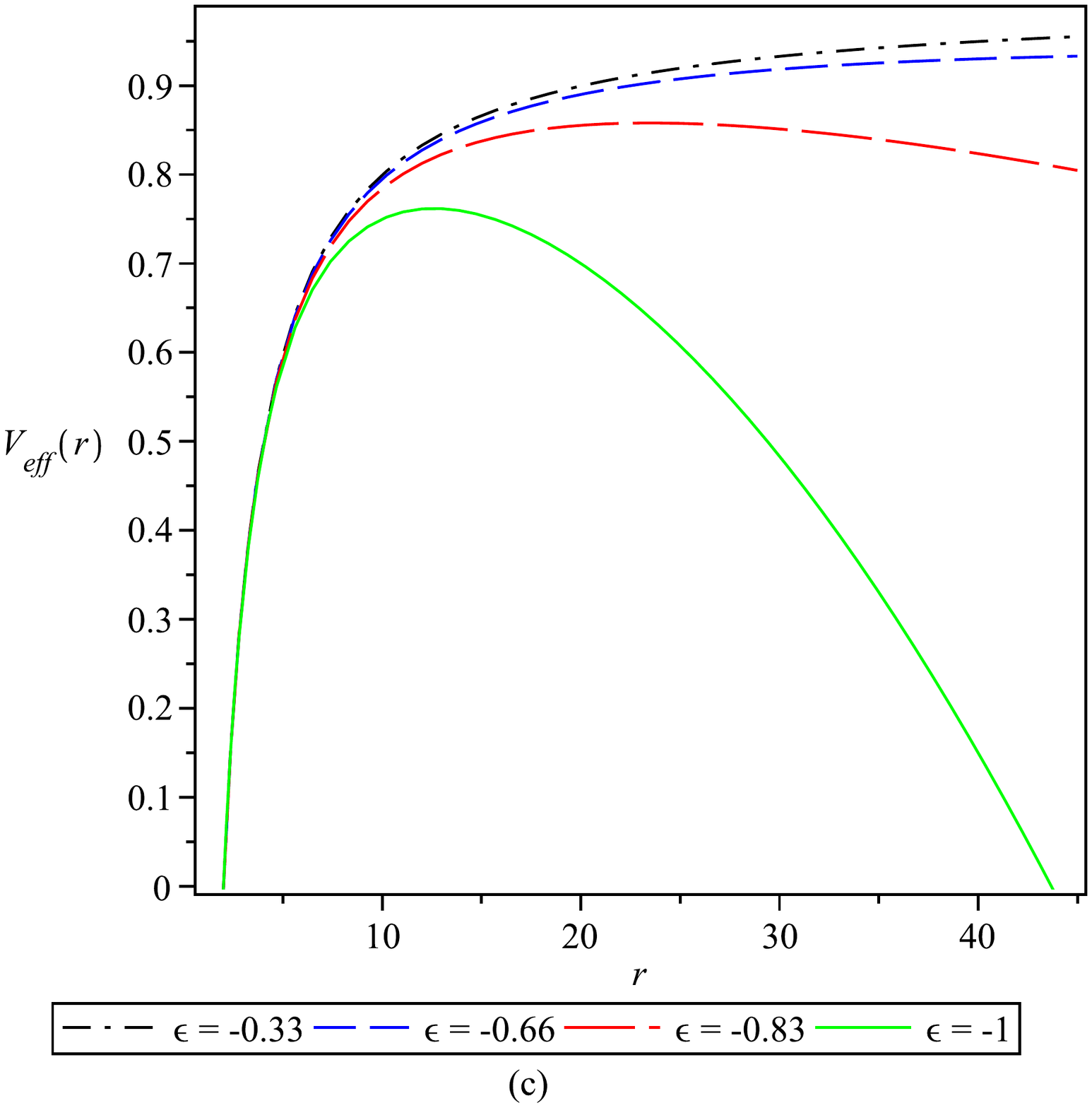}
\label{fig:p_9}
\caption{Effective potential for different values of equation of state parameter, $\epsilon$ (label in the figure), with a unit black hole mass and $\alpha=0.0005$ for different values of angular momentum,
$(a)$ ${L^2}=20$,
$(b)$ ${L^2}=9$ and
$(c)$ ${L^2}=0$.}
\label{fig:pp_2}
\end{figure}

Here the first three terms come out to be exactly same as that of the standard Schwarzschild case (first term represents the Newtonian gravitational potential, second term represents a repulsive centrifugal potential and third term as a relativistic correction of general relativity, i.e. proportional to $1/r^3$). The extra term $ \alpha(1+{L^2}/{r^2})/r^{3\epsilon+1}$ in eq. (\ref{potential}) is due to the presence of quintessence scalar field around the Schwarzschild black hole.
The changes coming out due to this extra term in the behaviour of the effective potential are shown in fig. (\ref{fig:pp_1}) and fig. (\ref{fig:pp_2}), for some specific values of $\epsilon$ (say $ \epsilon = -1/3, -2/3, -1$). More specifically, with $\epsilon = -1/3$, the last term, $\alpha(\frac{L^2}{r^2}+1)$ appears as a correction term in the centrifugal potential energy along with an extra attractive term, $-\alpha$ and with $\epsilon = -2/3$, the last term comes out as a correction term to the Newtonian gravitational potential, $-\alpha L^2/r$ with an attractive term of $-r\alpha $. Similarly, for $\epsilon = -1$, the correction terms appear as $-L^2\alpha $ and $-r^2\alpha$.
Hence, it is observed that the correction terms highly depend on the behaviour of scalar field through the equation of state.
 It is worth to mention at this moment that the equation of state value, $\epsilon = -1/3$,
an extreme limit for driving an accelerating expansion of our universe, is allowed neither by the current observation of SNe Ia nor by combined observations of CMB, BAO and
Hubble measurements. Indeed, the constrain from recent observational data of CMB and SNe Ia (Planck+WMAP9+Union2.1)\cite{Ade:2013zuv} is ${-1.16} < \epsilon < {-0.92}$ at the $95\%$ confidence level, favouring the dark energy model of phantom type with $\epsilon$ less than $-1$. However, our main motivation for this work is to study the geodesics and the behaviour of orbits of a test particle around a Schwarzschild black hole under the influence of a scalar field. Therefore, we restrict our calculation only to $-1 \leq \epsilon \leq -1/3$ and one can certainly consider the values beyond it and we hope to further investigate the other possibilities of dark energy model in the future work.
\vspace{-7.5mm}
\begin{figure}[hbtp]
\centering
\includegraphics*[width=5.6cm,height=4cm]{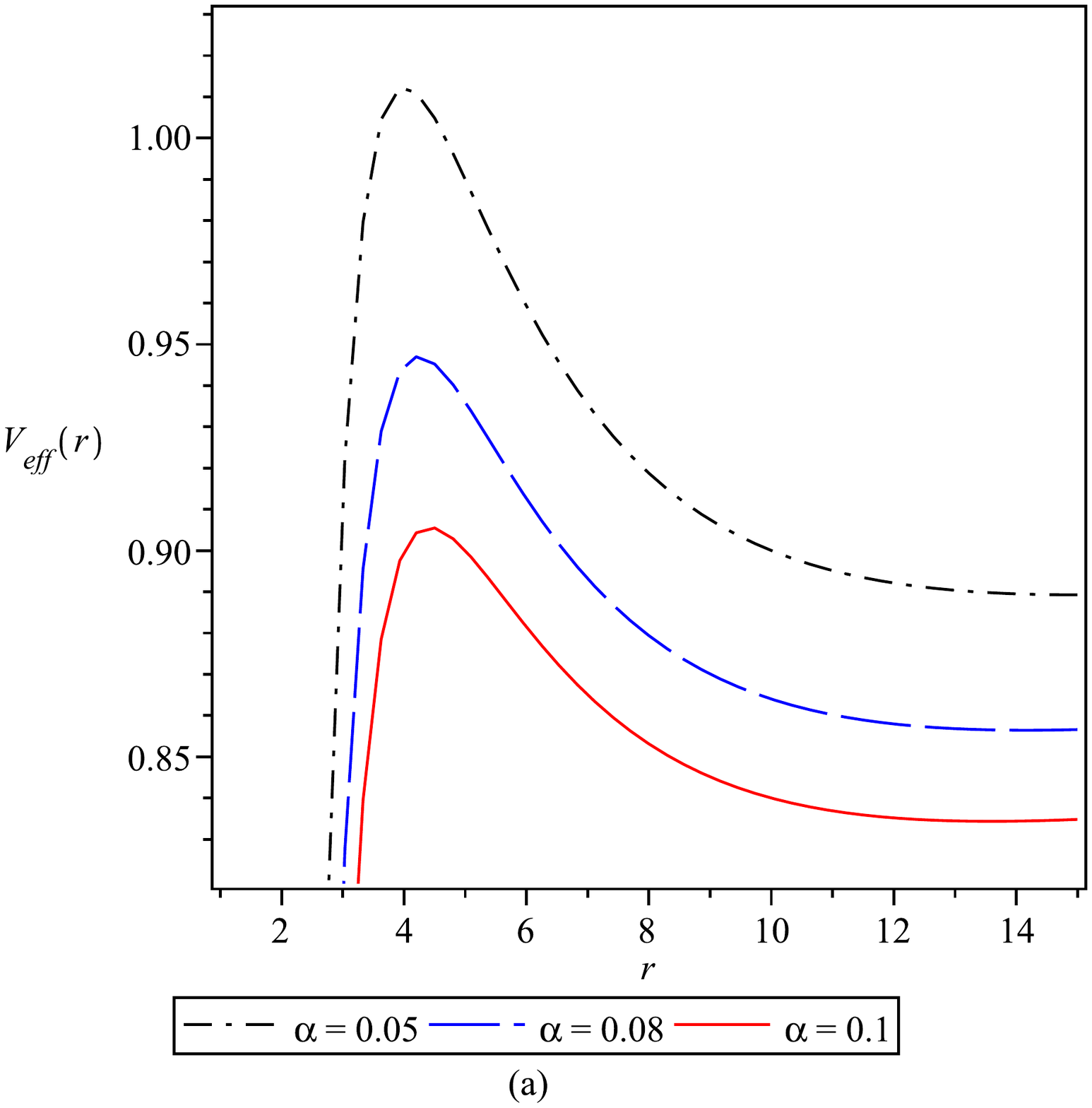}
\label{fig:p_1}
\hspace{-1mm}
\includegraphics*[width=5.6cm,height=4cm]{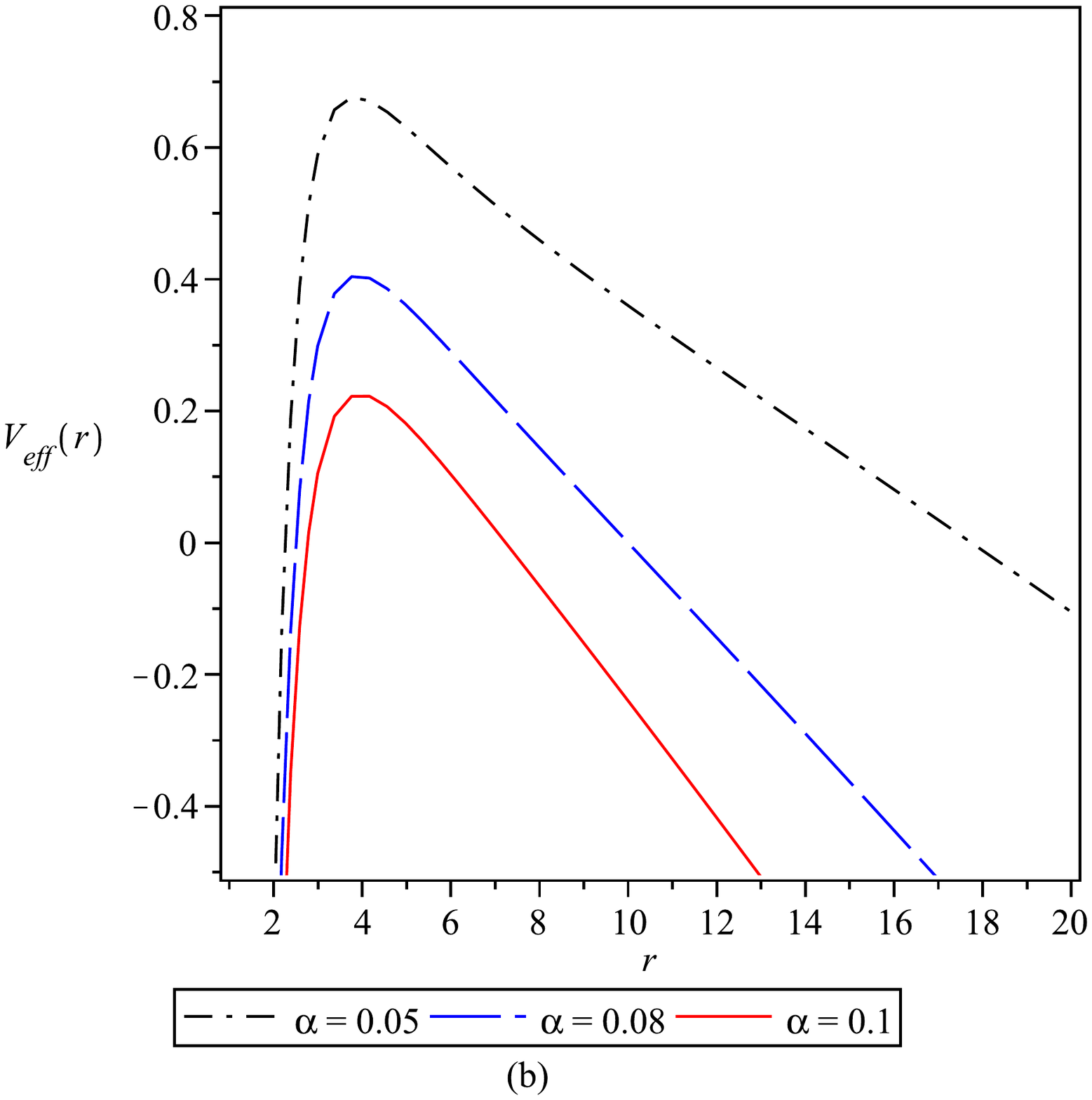}
\label{fig:p_2}
\nonumber
\end{figure}
\vspace{-10mm}
\section{\label{sec:level1}Nature of effective potential}
\begin{figure}[h]
\centering
\includegraphics*[width=8cm]{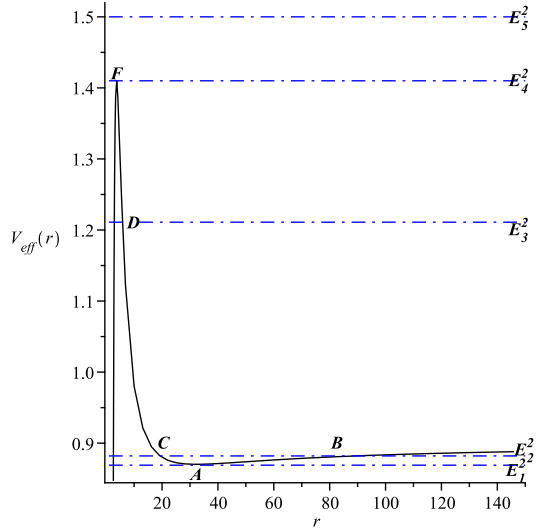}
\caption{Effective potential for a unit mass of black hole, angular momentum value of ${L^2} =40$, and equation of state parameter, $\epsilon=-1/3$, with the normalisation parameter, $\alpha = 0.1$.}
\label{fig:p}
\end{figure}

From eq. ({\ref{eqmotion}}) and eq. (\ref{potential}), one can discuss the time-like geodesics of different cases for both the radial ($L = 0$) and non-radial  geodesics ($L\neq0$). First, let us consider a specific case of non-radial geodesics with ${L^2}=40$ for a black hole of unit mass (i.e. $M=1$) and the equation of state parameter, $\epsilon = -1/3$ with $\alpha=0.1$, and try to understand the motion of a test particle along non radial time-like geodesics (see fig. (\ref{fig:p})) as follows:
\vspace{-0.2cm}
\begin{enumerate}
\item[(i)] {If $E>{E_4}$ (as the case of $E={E_5}$, where $E$ is the energy of incoming test particle) the particle will fall directly into the singularity starting from rest (a finite distance). In such cases we expect a plunge orbit in which the particle comes in from infinity, moves part way around the central mass and then plunges into the center.}
\item[(ii)] {If $E={E_4}$, the particle has an unstable circular orbit at point $F$ in the fig. (\ref{fig:p}), it may fall into the singularity beyond this, depending on initial energy conditions of the particle.}
\item[(iii)] {If $E>1-\alpha$ (as the case of $E={E_3}$), the particle will have a fly-by orbit, i.e. the particle comes from infinity, moves towards the center and after approaching a minimum distance (say point $D$ in the fig. (\ref{fig:p})), it flies again back towards the infinity.}
\item[(iv)]{If $E={E_2}$, the particle shows a bounded circular motion between points $B$ and $C$ which represent aphelion and perihelion distances respectively. }
\item[(v)]{If $E={E_1}$, with this energy particle strikes the minima of potential energy curve. This represents the possibility of bounded circular motion with radius equivalent to the distance of the minima (i.e. point $A$ in the fig. (\ref{fig:p})).}
\end{enumerate}

\subsection{\label{sec:level2}Analytic solution for the orbits of test particle}
It is well examined that universe exhibits an accelerated expansion \cite{Copeland:2006wr,Nozari:2008ff} for $0\leq3(1+\epsilon)<2$ and the values of $\epsilon$ lie within the range of $-1\leq\epsilon<-\frac{1}{3}$. Though, we have mentioned before,  
$\epsilon=-1/3$ is the border value for acceleration and deceleration \cite{Copeland:2006wr}, it however provides a simplified context for analytic solutions of orbit equations. 
Therefore, in view of the availability of such analytic solutions,
different types of orbits for test particles following timelike geodesic congruences for $\epsilon=-1/3$ are presented in this section.\\ In order to have a complete analytic description of different types of orbits, a careful attention is needed further and is a matter of separate discussion.
\begin{figure}[htbp]
\centering
\includegraphics*[width=4.5cm,height=4.0cm]{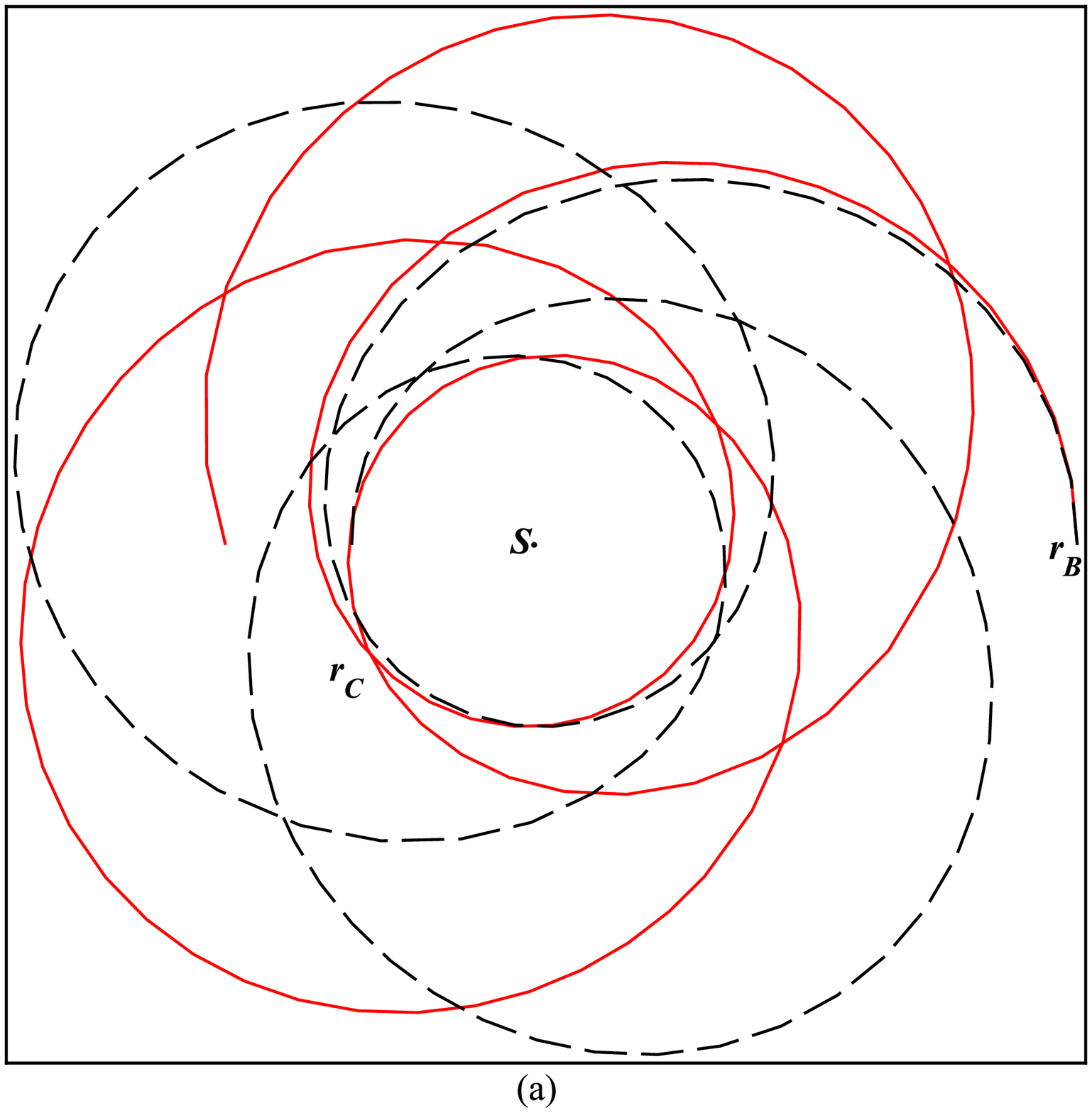}
\label{fig:p1}
\includegraphics*[width=6.5cm,height=4.0cm]{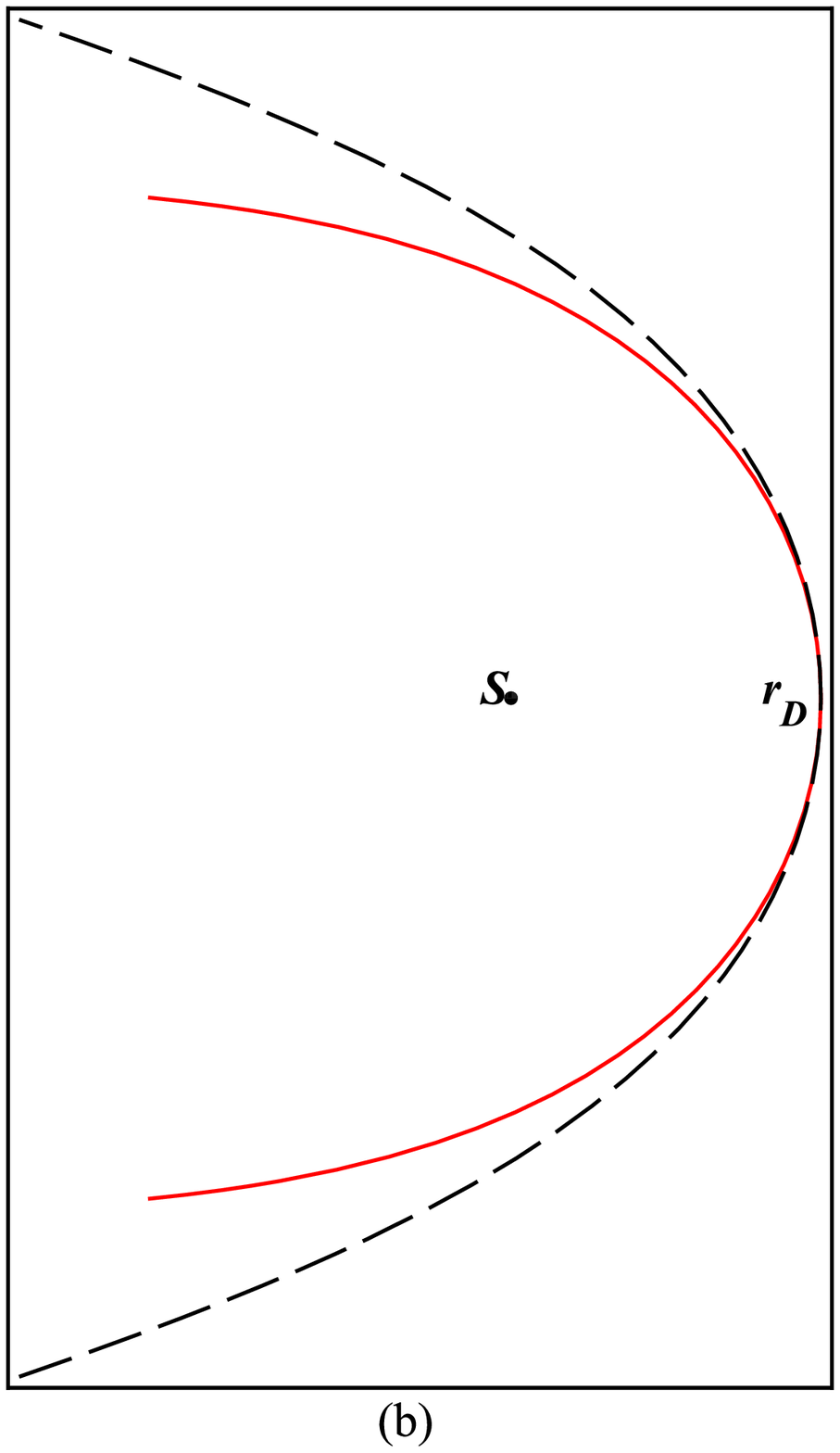}
\label{fig:p2}
\includegraphics*[width=4.5cm,height=4.1cm]{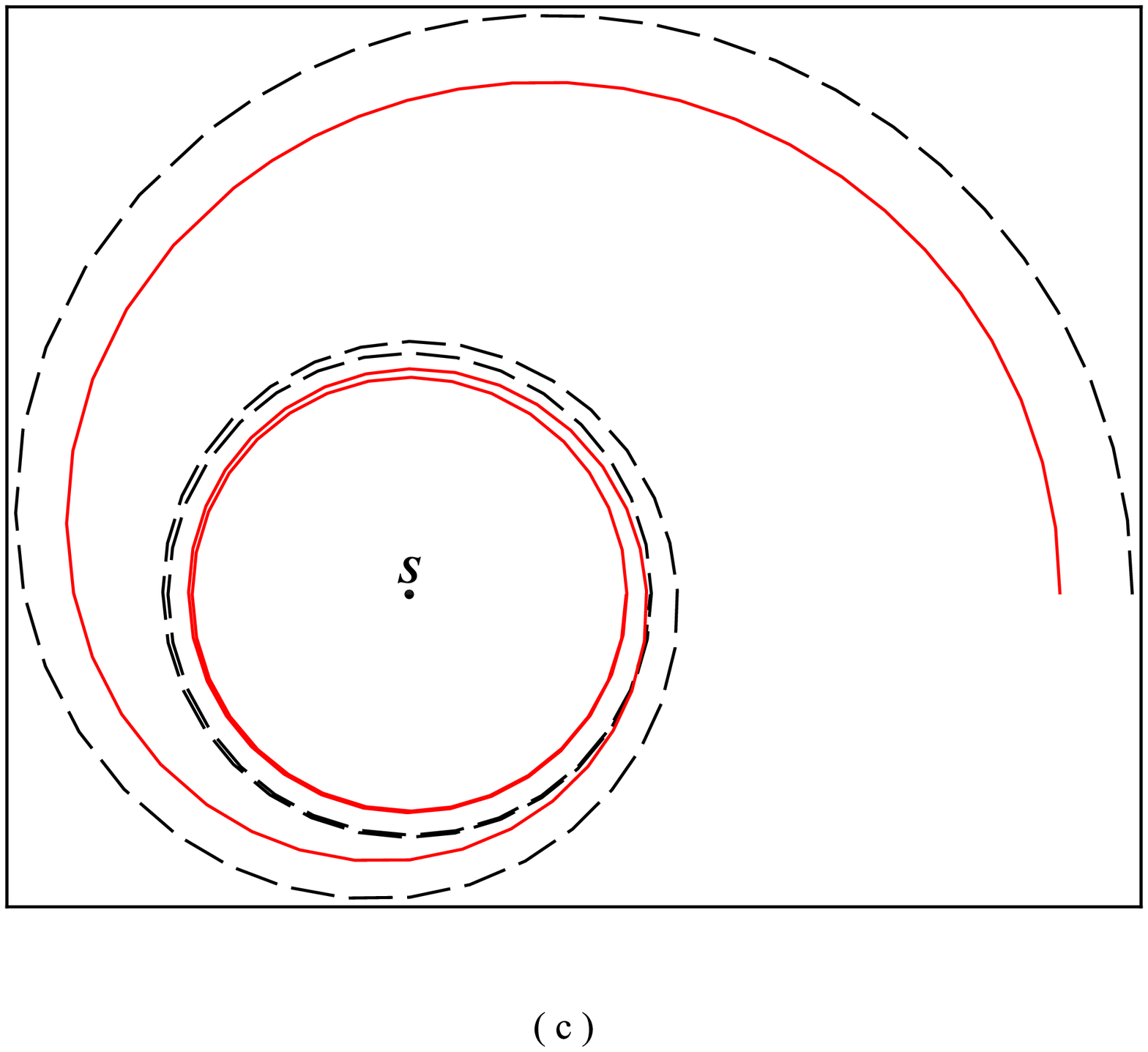}
\label{fig:p3}
\hspace{1cm}
\includegraphics*[width=4.5cm,height=4.5cm]{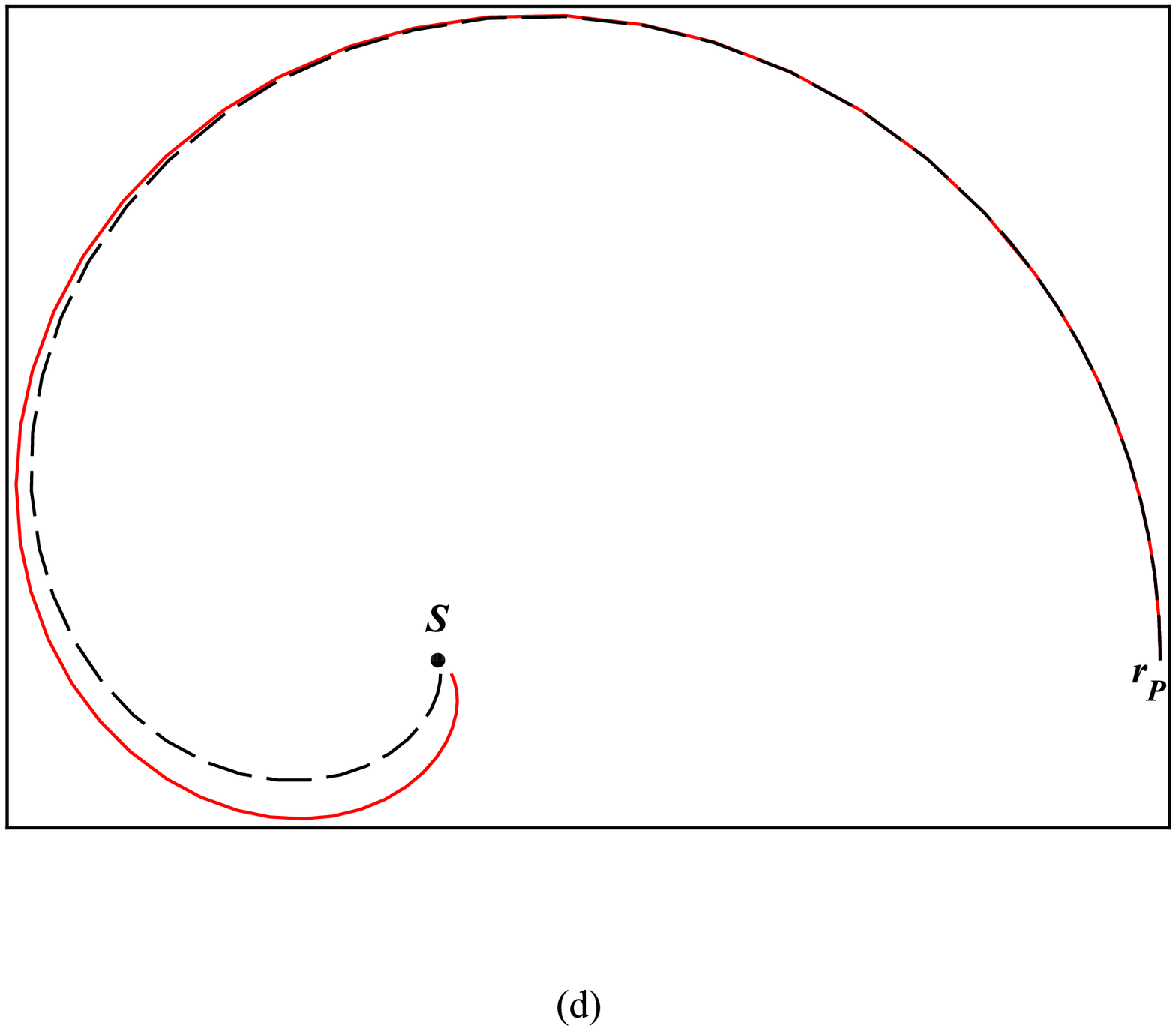}
\label{fig:p4}
\caption{Possible orbits of the regions described in fig. ({\ref{fig:p}}) for the same set of parameters (i.e. $M=1$, ${L^2}=40$, $\epsilon=-1/3$ and $\alpha=0.1$) but with different energies of an incoming test particle $E$:
(a) ${E^2}=0.882$,
(b) ${E^2}=1.211$,
(c) ${E^2}=1.41$ and
(d) ${E^2}=1.5$. Here, dotted and solid curves represent orbit for the test particle around Schwarzschild Black hole with quintessence and without quintessence field respectively, $S$ denotes the central singularity.}
\label{fig:orbits}
\end{figure}

In fig. (\ref{fig:orbits}), we present explicitly all the  possible orbits of the test particle corresponding to the discussions made above with different values of energy E.
\begin{enumerate}
\item[4(a).] Orbit for particle energy ${E}$ corresponding to ${E_2}( =0.882)$ in fig. ({\ref{fig:p}}), clearly shows a bound motion of the particle between points $B$ and $C$.
\item[4(b).] Orbit for particle energy ${E}$ corresponding to $E_3( = 1.211)$ in fig. ({\ref{fig:p}}), an existence of fly-by orbit with turning point at $D$ is displaying.
\item[4(c).] Orbit for particle energy $E$ corresponding to $E_4(=1.41)$ and we observe an unstable circular orbit at some point (say $F$ as in fig. ({\ref{fig:p}})).
\item[4(d).] Orbit for particle energy $E$ corresponding to $E>{E_4 (=1.5)}$ in fig. ({\ref{fig:p}}), as expected it shows the presence of a terminating orbit.
\end{enumerate}

\subsection{\label{sec:level2}\textbf{Circular and Innermost Stable Circular Orbits (ISCO)}}
For circular geodesics of constant $r$ and from eq.(\ref{eqmotion}) we have,
\begin{equation}
{V_{eff}}={E^2}
\label{eq:circular_condition_01}
\end{equation}
and
\begin{equation}
\frac{dV_{eff}}{dr}=0
\label{eq:circular_condition_02}
\end{equation}
Hence angular momentum per unit mass of the test particle along the circular orbit can be obtained by eq.(\ref{eq:circular_condition_02}) for respective value of the equation of state parameter $\epsilon$.\\
The most important class of orbits from astrophysical point of view are the innermost stable circular orbits \textbf(ISCO). 
These orbits occur at the point of inflection of the effective potential $V_{eff}$. 
Thus at the point of inflection
\begin{equation}
\frac{{d^2}V_{eff}}{d{r^2}}=0
\label{eq:ISCO_condition}
\end{equation}
\begin{figure}[ht]
\begin{center}
\includegraphics[scale=0.3]{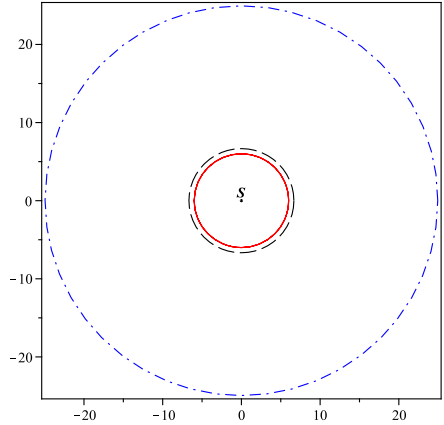}
\caption{The position of the innermost stable circular orbit with $M=1$ and $\alpha=0.1$ where the circle with solid line represents ISCO for Schwarzschild black hole without quintessence, circles with dashed and dot-dashed lines represent ISCO with equation of state parameter $\epsilon=-1/3,-2/3$ respectively and $S$ denotes the central singularity.}
\label{fig:circle}
\end{center}
\end{figure}
with auxiliary equation $\frac{dV_{eff}}{dr}=0$.
Hence ISCO equation for each case of representative $\epsilon$ value can be obtained from eq.(\ref{eq:circular_condition_02}) and eq.(\ref{eq:ISCO_condition}). 
Hence radius of ISCO in each case depends on parameters such as mass of the black hole $M$, angular momentum of test particle $L$ and normalization factor $\alpha$.
With $M=1$ and $\alpha=0.1$ real roots of ISCO equation are obtained only for $\epsilon=-1/3,-2/3$.
Hence it will be quite interesting to constraint $\alpha$ to obtain radius of ISCO for different values of $\epsilon$.
Analysis of extremum points for effective potential energy (i.e. $\frac{dV_{eff}}{dr}=0$) for $\epsilon =-1/3$ leads to the following condition for $r$,
\begin{equation}
r^2+\frac{L^2}{M}(\alpha-1)r+3L^2=0,
\label{epsilon13}
\end{equation}
which implies,
\begin{equation}
r_{min/max} = \frac{L^2(\alpha-1)}{2M}\left(-1 \pm \sqrt{1-\frac{12M^2}{(\alpha-1)^2 L^2}}\right).
\end{equation}
One can visualise easily from eq. (\ref{epsilon13}) that with $\alpha=1$, the extrema become imaginary numbers such that $\alpha$ cannot have unit value.
Therefore, the value of $\alpha$ is confined within $ 0< \alpha < 1$.
 However at $\alpha =0$, the standard Schwarzschild case is recovered automatically.
 The effective potential has one maximum and one minimum if  $\alpha < 0.01$ and $L/M > \sqrt{12}$. The maximum lies above $V_{eff} = 0$ if $\alpha < 0.005$ and $L/M > \sqrt{12}$ irrespective of $r$ value. Larger circular orbit locates the minima of potential energy, hence it corresponds to orbit of Ist kind, presenting a Stable Circular Orbit with radius $r_c$. It is shown in fig. (\ref{fig:circle}) that due to the presence of quintessence, the radius of innermost stable circular orbits is shifted to larger distance from center as compared to the case of Schwarzschild black hole without quintessence. However, for $\epsilon = -1/3$, the condition of innermost stable circular orbit, (ISCO) i.e. ${L^2}/{M^2}=12/{(1-\alpha)^2}$ can be obtained from the condition on stable circular radius ($r_c$) to be real, i.e. ${{L^2}/{(1-\alpha)^2}{M^2}}\geq 12$.
For $\alpha = 0$, this condition for ISCO reduces to the pure Schwarzschild case.

\subsection{\label{sec:level2}Numeric visualization of orbits }
Despite the absence of corresponding exact analytic solutions of orbit equations, the numeric solutions for the values of $\epsilon$ other than $\epsilon=-1/3$ in the range $-1\leq\epsilon<-1/3$ are presented in this section.

\begin{figure}[ht]
\begin{tabular}{cc}
\includegraphics[width=4.5 cm, height=6 cm]{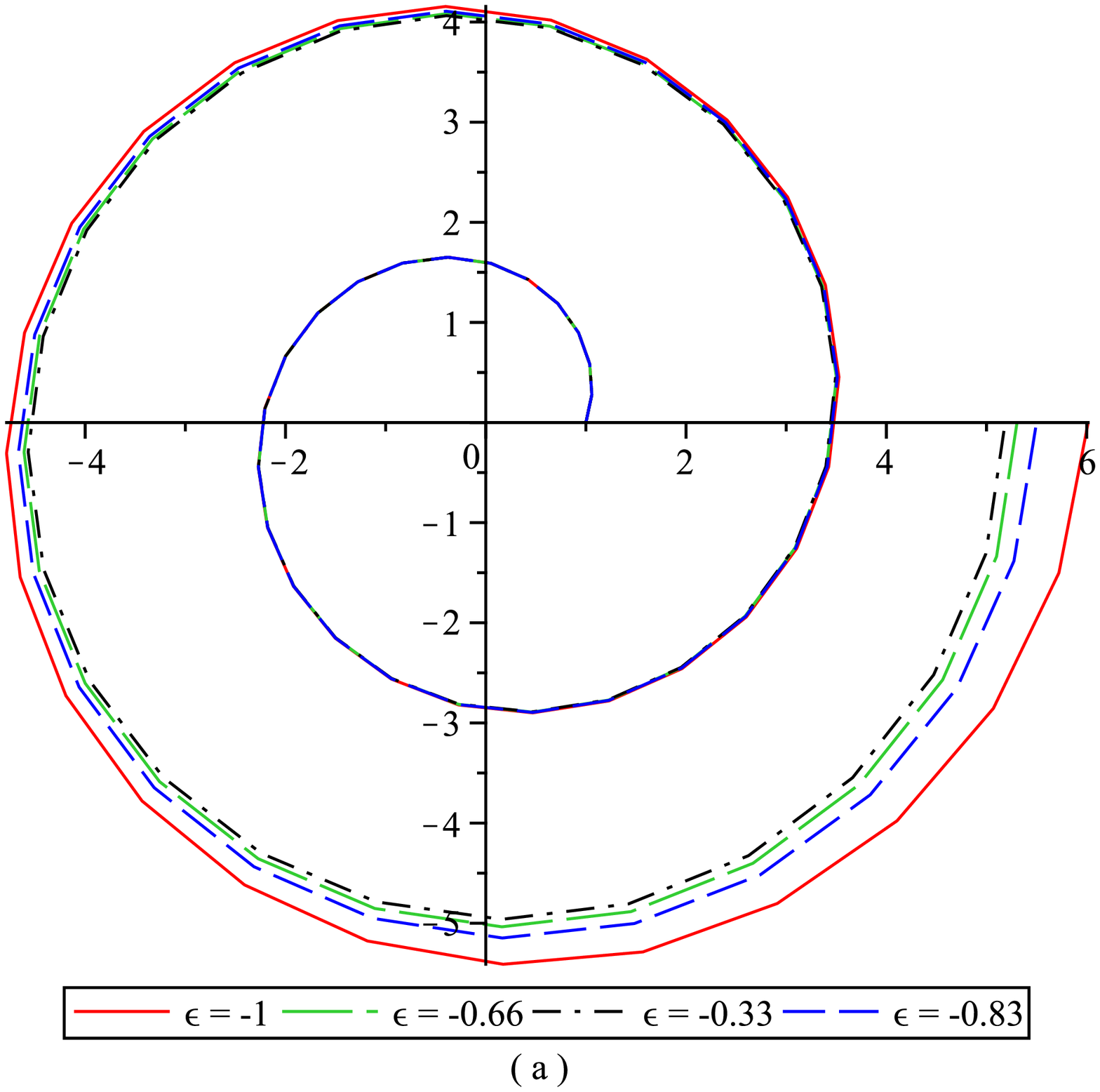}&
\includegraphics[width=5 cm, height=6 cm]{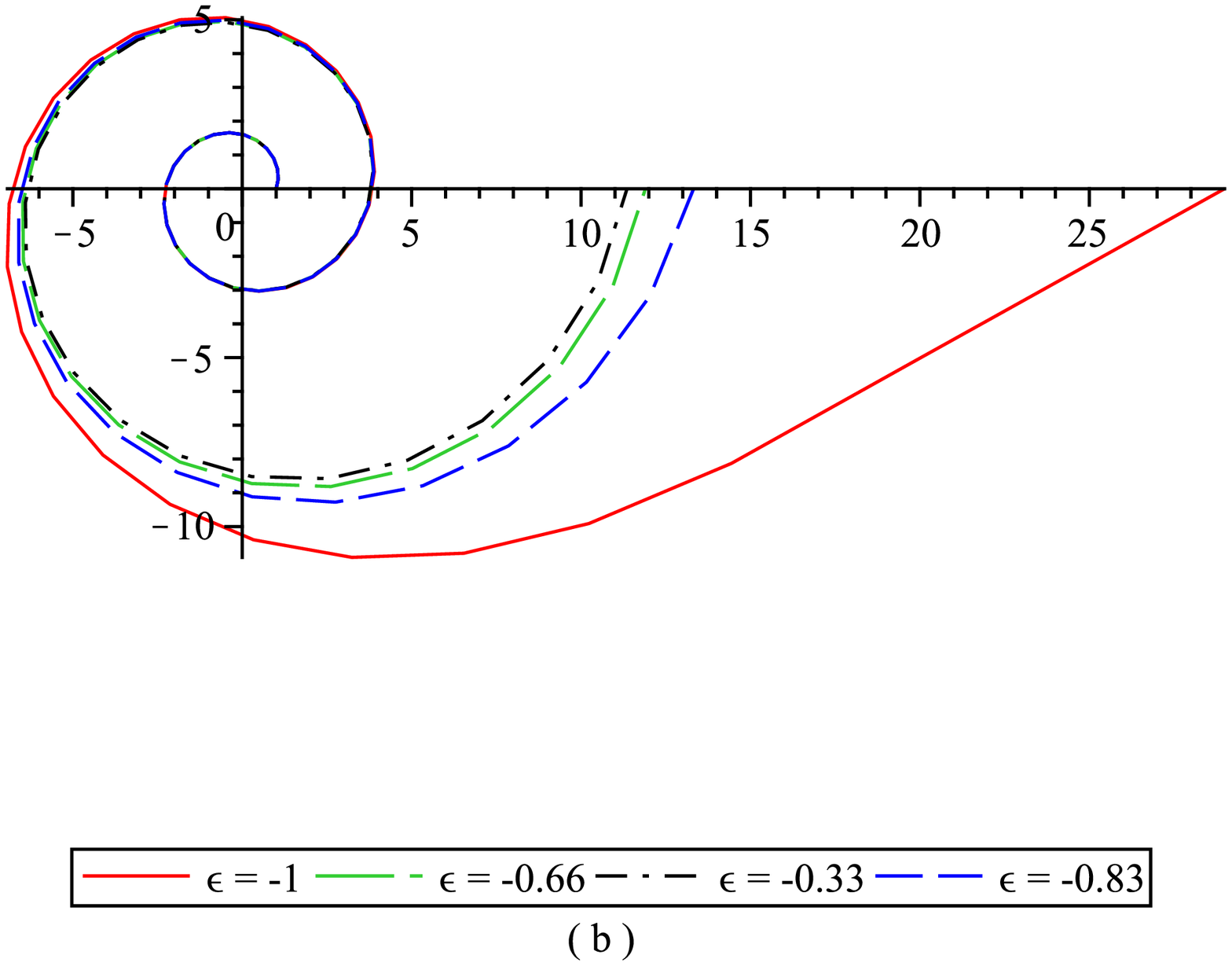}
\\
\hline
\\
\includegraphics[width=4.5 cm, height=6 cm]{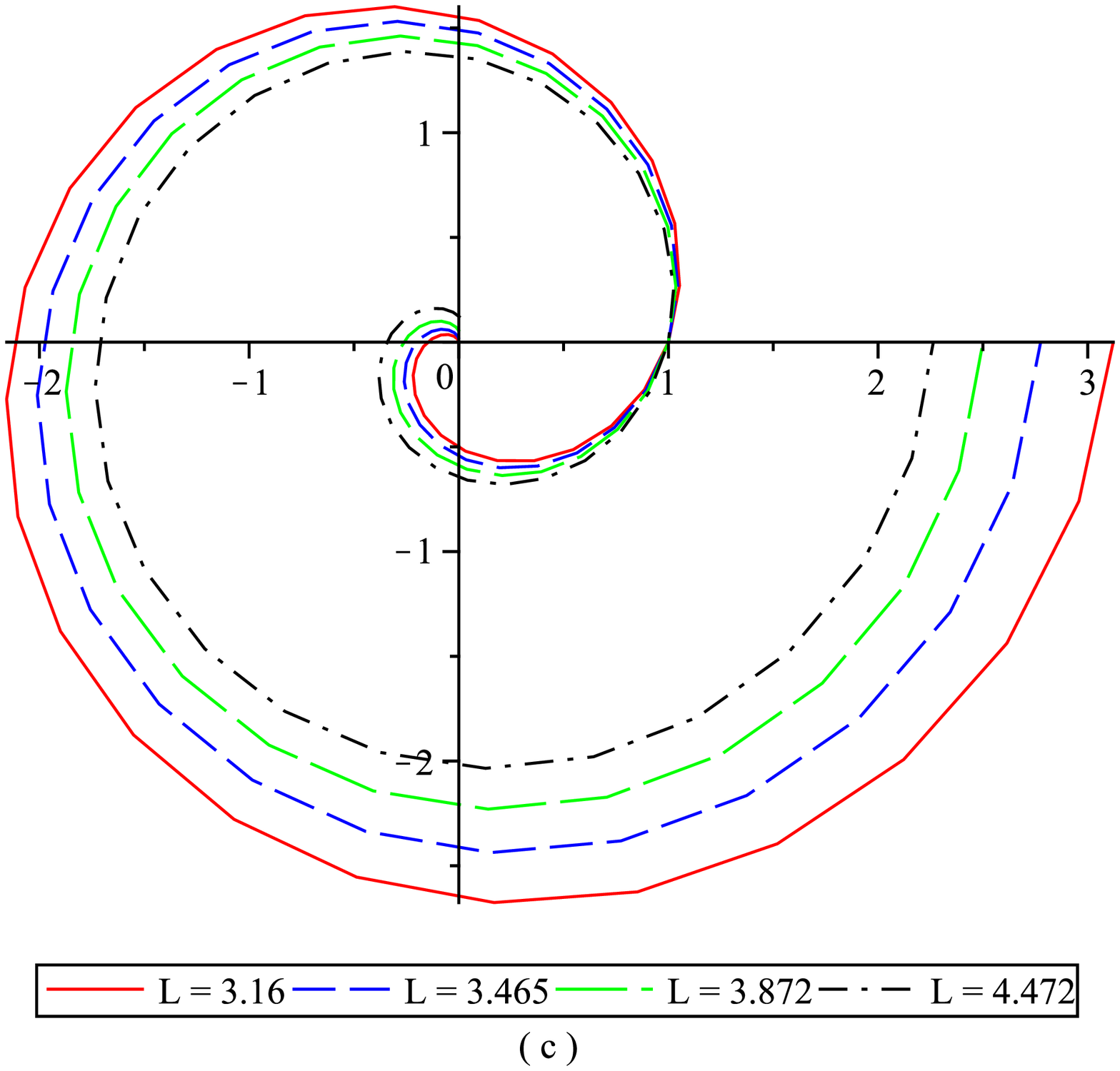}&
\includegraphics[width=4.5 cm, height=6 cm]{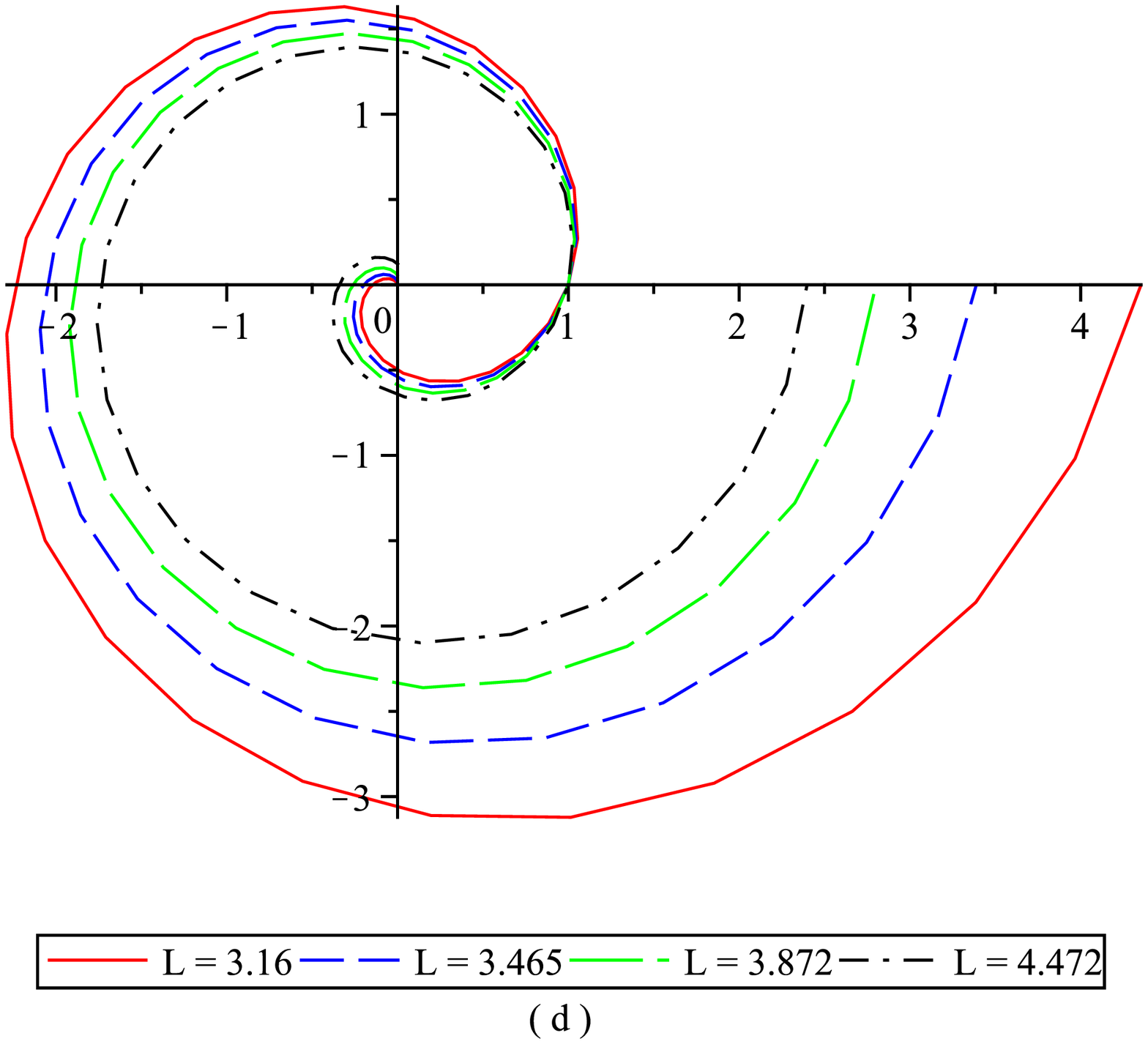}
\\
\end{tabular}
\caption{{\it Upper Panels} -- Orbits of test particles for various values of $\epsilon$ (see labels in the figure) with $L=3$ at two sightly different energies : (a) ${E^2}=0.82$ and (b) ${E^2}=0.9$. {\it Lower Panels} -- the same as upper panels but with different values of $L$ for $\epsilon=-1/3$ :
(c) ${E^2}=0.8$ and
(d) ${E^2}=1.15$. Here $\alpha=0.0005$ and $r(0)=1$.}
\label{fig:numeric_orbit}
\end{figure}

From fig. (\ref{fig:numeric_orbit}), one can mark that how do the orbits of test particles behave with varying quintessence parameter, $\epsilon$ (or angular momentum, $L$) for particular values of energy $E$ and other parameters.
 Now, the effect of quintessence parameter can be observed directly from the fig. (\ref{fig:numeric_orbit}a) and fig. (\ref{fig:numeric_orbit}b) with the decrease in $\epsilon$ from $-1/3$ to $-1$, particles start circulating towards the center from larger radial distances. In other words, smaller the $\epsilon$ value, larger the radial distance it has. However, the orbits are quite similar for the values of $\epsilon$ we considered, in the vicinity of center. This is clearly shown in fig. (\ref{fig:numeric_orbit}a) and fig. (\ref{fig:numeric_orbit}b).

On analysing the orbit for test particles at some particular energy $E$ but for various angular
momentum $L$ values (refer to fig. (\ref{fig:numeric_orbit}c) and fig. (\ref{fig:numeric_orbit}d)), it is found that particles with smaller $L$ values are set into circular motion around black hole from larger distances.\\
\\
In order to understand the possible orbits for test particle (having energy $E$), along radial time-like geodesics ($L=0$), we again consider a specific case of  $M=1$,  equation of state parameter $\epsilon=-5/6$ and $\alpha=0.0005$. The result is shown in fig. (\ref{fig:p_radial}) and discussed below:
\begin{figure}[ht]
\centering
\includegraphics*[width=8cm]{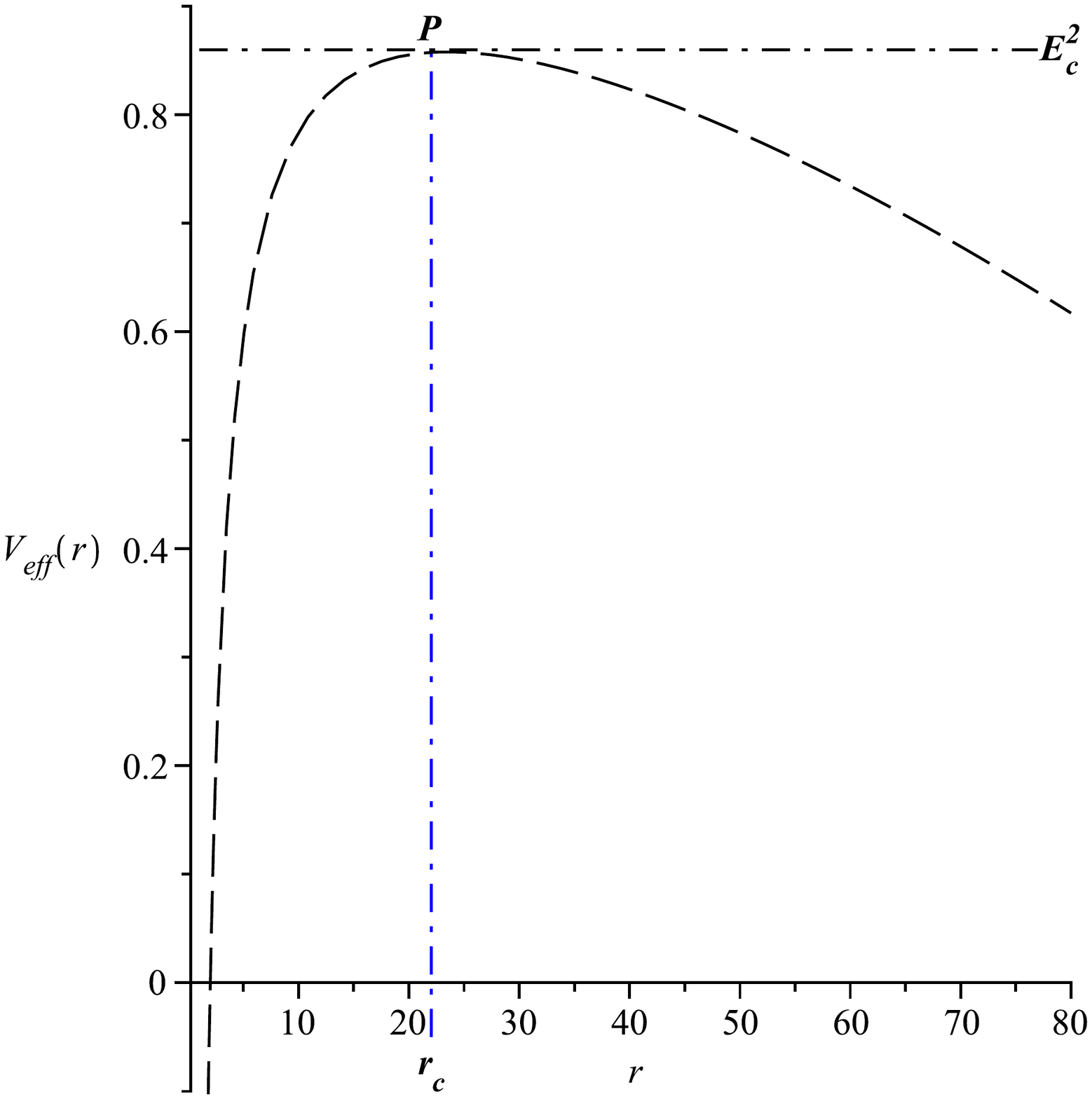}
\caption{Effective potential for the radial geodesics (${L} =0$) with $M=1$, $\epsilon=-5/6$ and  $\alpha=0.0005$.}
\label{fig:p_radial}
\end{figure}
\begin{enumerate}
\item {If $E>{E_c}$ (see fig. (\ref{fig:p_radial})), the particle will have only a kind of terminating orbit, i.e the particle will start form infinity and finally  will drop into a singularity.}
\item {If $E={E_c}$, the particle will have an unstable circular orbit at this point (say $P$ in fig. (\ref{fig:p_radial}))  with a radius $r = {r_c}$.}
\item {If $E<{E_c}$, the particle will either plunge into the singularity for the radii $r<{r_c}$ or fly back to the infinity for $r>{r_c}$. }
\end{enumerate}

Therefore, no bound orbits are possible for radial geodesics. The only possible orbits are either terminating or fly by orbits.

\section{\label{sec:level1}Geodesic Deviation}
Another interesting way to observe the effect of quintessence field is through the equation of geodesic deviation, which depicts the relative acceleration of some test particles falling freely in the gravitational field of a black hole.
 The study of geodesic deviation not only enables to understand the physical effects of the gravitational field but one can also get a clear idea about the effect of quintessence field on the geometry of the space time.
 We follow the method of \cite{Gad:2007qt}, in order to derive the equation of geodesic deviation (Jacobi field equation)
 
\begin{equation}
{\frac{{{D^2}{{\eta}^a}}}{D{{\tau}^2}}}+{{R^a}_{bcd}}{v^b}{v^c}{\eta^d}=0
\end{equation}

where $v^a$ represents a vector tangent to geodesic and $\eta^a$, a vector which shows the connection between two neighbouring geodesics.
Generally, the Jacobi field equation is studied by considering a congruence of time-like geodesics with a unit tangent vector  $v$($g(v,v)=-1$) and defining dual bases ($e^a_0$,~$e^a_1$,~$e^a_2$,~$e^a_3$)
and ($e^0_a$,~$e^1_a$,~$e^2_a$,~$e^3_a$) of the tangent space ${T_q}M$ and dual tangent ${T^*_q}M$ respectively, at some point $p$ on the geodesic  $\gamma$($\tau$). Letting the basis to be propagated along the time-like geodesic i.e. $e^a_0 = v^a$ and $e^1_a$,$e^2_a$,$e^3_a$ orthogonal to $v^a$ (see for more details\cite{Gad:2007qt}) and the orthogonal connecting vector, $\eta^a$ can be expressed as $\eta^a = \eta^\alpha e^a_{\alpha}$ and $\eta^0 = e^0_{\alpha}\eta^\alpha = 0$, connecting two neighbouring particles in free fall. Here $\eta^\alpha = (\eta^r,~\eta^\theta,~\eta^\phi)$ are the space-like components of $\eta^a$. The equation of geodesic deviation vector $\eta^a$ reduces to

\begin{equation}
{\frac{{{D^2}{{\eta}^\alpha}}}{D{{\tau}^2}}}+{\tilde{R}}^a_{bdc}{e^\alpha_a}{v^b}{v^c}{e^d_\beta}{\eta^\beta}=0
\label{eq:devaition2}
\end{equation}

where the Riemann tensor ${\tilde{R}}^a_{bdc}$ is written in the frame of ($e^a_0$,~$e^a_1$,~$e^a_2$,~$e^a_3$).
For our case of Schwarzschild black-hole surrounded by the quintessence field, the frame $e^a_b$ takes the form:
\begin{equation}
{e^a_0}=[f(r)]^{\frac{-1}{2}}(0,0,0,1);~
{e^a_1}=[f(r)]^{\frac{1}{2}}(1,0,0,0);~
\nonumber
\end{equation}
\begin{equation}
{e^a_2}={\frac{1}{r}}(0,1,0,0);~
{e^a_3}={\frac{1}{r\sin\theta}}(0,0,1,0).
\label{eq:eframe}
\end{equation}

Using ${v^a}={e^a_0}$, Riemann tensor of spacetime given in eq. (\ref{lineelement}), eq. (\ref{eq:eframe}) and converting the covariant derivative into ordinary derivatives, ${\frac{D\eta^\alpha}{D\tau}}={\frac{d\eta^\alpha}{d\tau}}+
{\tilde{\Gamma}^\alpha_{ab}}{\eta^b}{v^a}$ ,
the eq. (\ref{eq:devaition2}) becomes (in the components form):

\begin{equation}
{\frac{{{d^2}{{\eta}^r}}}{d{{\tau}^2}}}-\left[\frac{\alpha(2+9\epsilon+9{\epsilon}^2)}{2 r^{3\epsilon+3}}+\frac{2M}{r^{3}}\right]{\eta^r}=0,\label{eq:eta_r1}
\end{equation}
\begin{equation}
{\frac{{{d^2}{{\eta}^\theta}}}{d{{\tau}^2}}}+\left[\frac{\alpha(1+3\epsilon)}{2r^{3\epsilon+3}}+\frac{M}{r^3}\right]{\eta^\theta}=0,\label{eq:eta_th1}
\end{equation}
\begin{equation}
{\frac{{{d^2}{{\eta}^\phi}}}{d{{\tau}^2}}}+\left[\frac{\alpha(1+3\epsilon)}{2r^{3\epsilon+3}}+\frac{M}{r^3}\right]{\eta^\phi}=0.\label{eq:eta_ph1}
\end{equation}

As usual, eq. (\ref{eq:eta_r1}) represents the tidal force effect in radial direction while the eq. (\ref{eq:eta_th1}) and eq. (\ref{eq:eta_ph1}) manifest the pressure or compression effects in the transverse directions.
The standard geodesic deviation equations for the Schwarzschild black hole can be recovered once $\alpha$ is set to zero.
For the case of freely falling particles with zero angular momentum (L=0), the relation between
radial coordinate $r$ and the affine parameter $\tau$ can be obtained easily as

\begin{equation}
\frac{dr}{d\tau}=-\sqrt{({E^2}-1)+\frac{2M}{r}+\frac{\alpha}{2r^{3\epsilon+1}}}\label{eq:r_tau}
\end{equation}

With eq. (\ref{eq:r_tau}), one can rewrite the system of geodesics deviation eqs. (4.4), in term of radial coordinate derivative such as:

\begin{equation}
{\frac{{d^2}{\eta^r}}{d{r^2}}}
-\frac{1}{2r}\left[\frac{\frac{2M}{r}+\frac{\alpha(3\epsilon+1)}{r^{3\epsilon+1}}}{(E^2-1)+\frac{2M}{r}+\frac{\alpha}{r^{3\epsilon+1}}}\right]
{\frac{{d}{\eta^r}}{d{r}}}
-\frac{1}{r^2}\left[\frac{\frac{\alpha(2+9\epsilon+9{\epsilon}^2)}{2r^{3\epsilon+1}}+\frac{2M}{r}}{(E^2-1)
+\frac{2M}{r}+\frac{\alpha}{r^{3\epsilon+1}}}\right]{\eta^r}=0
\end{equation}
\begin{equation}
{\frac{{d^2}{\eta^{\theta}}}{d{r^2}}}
-\frac{1}{2r}\left[\frac{\frac{2M}{r}+\frac{\alpha(3\epsilon+1)}{r^{3\epsilon+1}}}{(E^2-1)+\frac{2M}{r}+\frac{\alpha}{r^{3\epsilon+1}}}\right]
{\frac{{d}{\eta^{\theta}}}{d{r}}}
+\frac{1}{2r^2}\left[\frac{\frac{2M}{r}+\frac{\alpha(1+3\epsilon)}{r^{3\epsilon+1}}}{(E^2-1)+\frac{2M}{r}+\frac{\alpha}{r^{3\epsilon+1}}}\right]{\eta^{\theta}}=0
\end{equation}
\begin{equation}
{\frac{{d^2}{\eta^{\phi}}}{d{r^2}}}
-\frac{1}{2r}\left[\frac{\frac{2M}{r}+\frac{\alpha(3\epsilon+1)}{r^{3\epsilon+1}}}{(E^2-1)+\frac{2M}{r}+\frac{\alpha}{r^{3\epsilon+1}}}\right]
{\frac{{d}{\eta^{\phi}}}{d{r}}}
+\frac{1}{2r^2}\left[\frac{\frac{2M}{r}+\frac{\alpha(1+3\epsilon)}{r^{3\epsilon+1}}}{(E^2-1)+\frac{2M}{r}+\frac{\alpha}{r^{3\epsilon+1}}}\right]{\eta^{\phi}}=0.
\end{equation}
\label{eq:devaition-general}
 We further study the above equations, providing some specific values of $\epsilon $ as considered before (i.e. $\epsilon = -1/3, -2/3 ,-1$).\\

\noindent{ \textbf{Case I}: $\epsilon={-\frac{1}{3}}$}

For $\epsilon={-\frac{1}{3}}$ and with ${E^2} = 1-\alpha$, deviation eqs. (\ref{eq:devaition-general}) exactly reduce to the form of the standard Schwarzschild case with the energy difference of $\alpha$ factor.

The corresponding solutions of geodesic equations are given by:

\begin{equation}
\eta^{r}(r) =\frac{C_1}{\sqrt{r}}+{C_2}r~~{\rm and}~~
\eta^{\theta}(r) = \eta^{\phi}(r) = {C_3}\sqrt{r} + C_{4}r
\end{equation}
where $C_1$,~$C_2$,~$C_3$ and $C_4$ are constants of integration.
Instead of ${E^2} = 1-\alpha$, if we consider $E^2 =1$, then the solutions of deviation equations take the following forms:

\begin{equation}
\eta^{r}(r) ={D_1}{\left[{\frac{2M}{r}+\alpha }\right]}^{1/2}
+ D_{2} \left(\frac{6M}{\alpha^{2}}+\frac{r}{\alpha} - \frac{6M}{\alpha^{5/2}}{\left[{\frac{2M}{r}+\alpha}\right]}^{1/2}\right)
{\log}~\left[2(\alpha\sqrt{r}+\sqrt{a}\sqrt{2M+\alpha r})\right]
\end{equation}
\begin{equation}
\eta^{\theta}(r) = \eta^{\phi}(r) = {D_3} \,r - D_4\left(\frac{\sqrt{2Mr + \alpha {r^2}}}{M}\right)
\end{equation}

with a new set of constant of integrations $D_1$,~$D_2$,~$D_3$ and $D_4$.
Here we found that along with the similar dependency on $r$ of the standard Schwarzschild solutions, there are many more extra terms appeared due to the presence of quintessence field around the black-hole. The consequences of these additional terms will discuss at the end of this section.\\

\noindent{ \textbf{Case II}: $\epsilon = -\frac{2}{3}$}\\
As it is difficult to obtain an exact solution for geodesic deviation eqs. (\ref{eq:devaition-general}) for $\epsilon=-\frac{2}{3}$ case. Therefore, we restrict ourselves to a particular choice, ${\alpha}r=1-{\frac{2M}{r}}$, which represents the condition of horizon for this specific value of $\epsilon$.
Then, the generalized geodesic deviation equations reduce to the following set of equations (with ${E^2}=1$),

\begin{equation}
{\frac {d^{2}{\eta}^{r}}{d{r}^{2}}} -{\frac{(1-2\alpha r)}{2r}}{\frac {d{\eta}^{r}}{dr}}-\,
{\frac{2M}{{r^3}}}{\eta}^{r}=0
\end{equation}
\begin{equation}
{\frac {d^{2}{\eta}^{\theta}}{d{r}^{2}}} -{\frac{(1-2\alpha r)}{2}}\,
{\frac {d{\eta}^{\theta}}{dr}}+\,
{\frac{(1-2\alpha r)}{2{r^2}}}{\eta}^{\theta}=0
\end{equation}
\begin{equation}
{\frac {d^{2}{\eta}^{\phi}}{d{r}^{2}}} -{\frac{(1-2\alpha r)}{2}}\,
{\frac {d{\eta}^{\phi}}{dr}}+\,
{\frac{(1-2\alpha r)}{2{r^2}}}{\eta}^{\phi}=0,
\end{equation}

and corresponding components of deviation vectors are come out to be:

\begin{equation}
{\eta^r}(r)= {B_1}\left(3 \,+\,{\frac{4M}{r}}\right) + 
{B_2}\left[\left({\sqrt{r}}\,+\,{\frac{2M}{\sqrt{r}}}\right)\, \exp\left(\frac{-2M}{r}\right) +
{\sqrt{\frac{M\pi}{2}}}\, \left(3+ \,{\frac{4M}{r}}\right)\,
 \rm erf\left(\sqrt{\frac{2M}{r}}\right)\right]
\end{equation}
\begin{equation}
{\eta^{\theta}} \left( r \right) = {\eta^{\phi}} \left( r \right) ={B_3}\,r \,+ {B_4}\,
\left[ r\sqrt{\alpha\pi}\, \left[\rm erf\left(\sqrt{\alpha r}\right)-1\right]\,
 +{\sqrt{r}} \exp \left(-\alpha r\right)\right]
\end{equation}

where $B_1$,~$B_2$,~$B_3$ and $B_4$ are integration constants and
$\rm erf(x)$ $=$
${\frac{2}{\sqrt\pi}}$ ${{\int}_0^x}{\exp\left(-{t^2}\right)}\, dt$.

\noindent{ \textbf{Case III}: $\epsilon=-1$}\\

Similar to the case of  $\epsilon = -\frac{2}{3}$,  for $\epsilon=-1$, we again rewrite the system of deviation equations with a new horizon condition, $\alpha {r^2} = 1-2M/r$ into the following forms:

\begin{equation}
{\frac{{d^2}{\eta^r}}{d{r^2}}}-\frac{1}{2r E^2}(1-3\alpha r^2)\frac{{d}{\eta^r}}{d{r}}-\frac{\eta^r}{r^2 E^2}=0
\end{equation}
\begin{equation}
\frac{{d^2}{\eta^{\theta}}}{d{r^2}}-\frac{1}{2r E^2}(1-3\alpha r^2)\frac{{d}{\eta^{\theta}}}{d{r}}+
 \frac{1}{2r^2 E^2}(1-3\alpha r^2){\eta^{\theta}}=0
\end{equation}
\begin{equation}
\frac{{d^2}{\eta^{\phi}}}{d{r^2}}-\frac{1}{2r{E^2}}(1-3\alpha r^2){\frac{{d}{\eta^{\phi}}}{d{r}}}+
\frac{1}{2r^2 E^2}(1-3\alpha r^2){\eta^{\phi}}=0
\end{equation}
\label{eq:e-1}

and give the solutions with $E^2=1$ as:

\begin{equation}
\eta^r(r) =\frac{1}{18}\exp\left(-\frac{3\alpha r^2}{4}\right)
\left[\frac{18A_1}{\sqrt{r}}+\frac{A_2}{\alpha}\left(12 \exp\left(\frac{3\alpha r^2}{4}\right)+\frac{\sqrt{2}~3^{3/4}}{(-\alpha r^2)^{1/4}}{\Gamma}\left [\frac{1}{4},-\frac{3\alpha r^2}{4}\right]\right)\right]
\end{equation}
\begin{equation}
\eta^{\theta}(r)= \eta^{\phi}(r)= {A_3} r + 
{A_4}\sqrt{r}\left(-2\exp\left(-\frac{3\alpha r^2}{4}\right)+\sqrt{2}~(3\alpha r^2)^{1/4}{\Gamma}\left[ \frac{3}{4},\frac{3\alpha r^2}{4}\right]\right)
\end{equation}

where $A_1$,~$A_2$,~$A_3$ and $A_4$ are integration constants.
\begin{figure}[hbtp]
\centering
\includegraphics*[width=8cm,height=6.1cm]{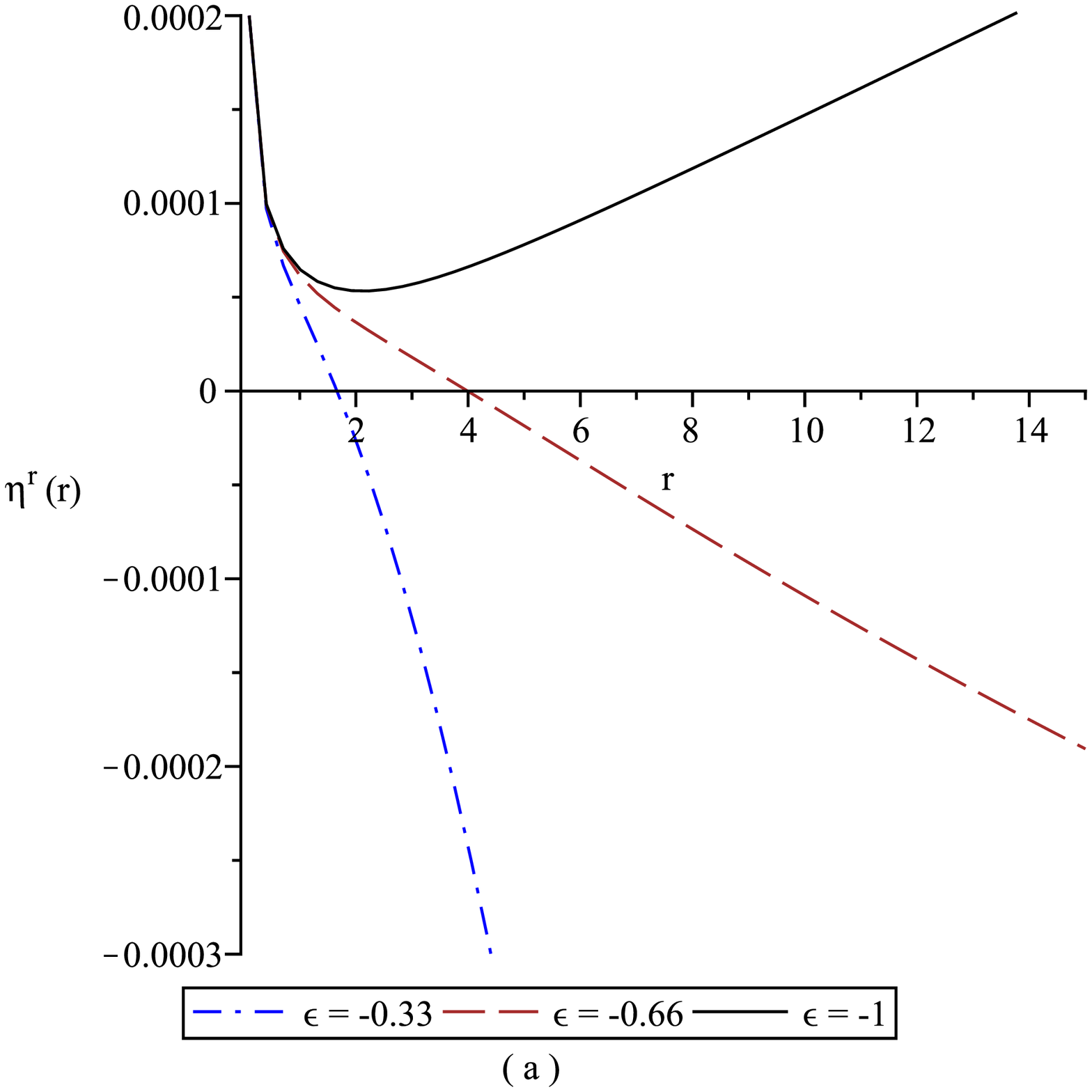}
\label{d1}
\nonumber
\end{figure}
\begin{figure}[hbtp]
\centering
\includegraphics*[width=8cm,height=6.1cm]{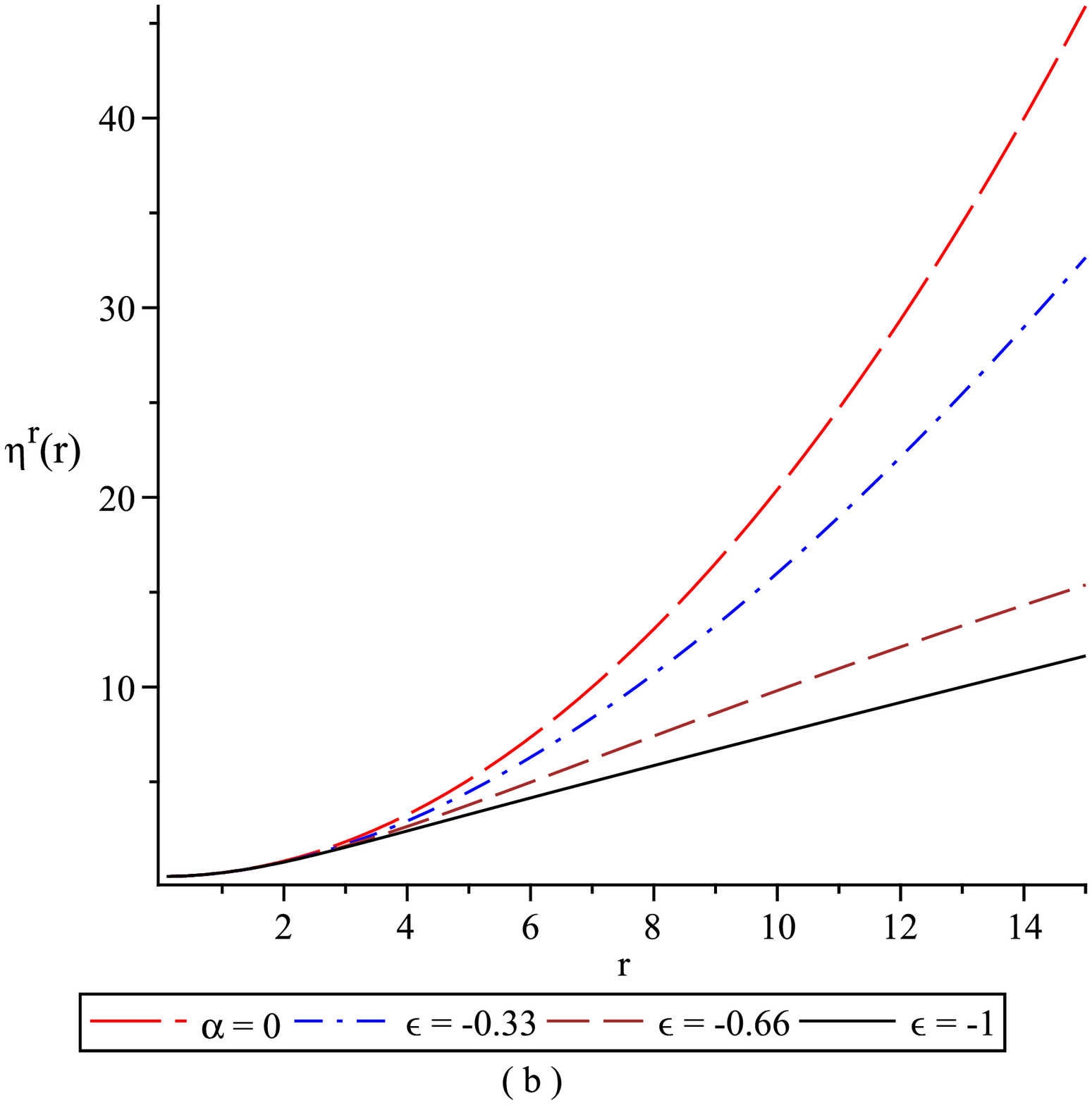}
\label{d2}
\nonumber
\end{figure}
\begin{figure}[hbtp]
\centering
\includegraphics*[width=8cm,height=6.1cm]{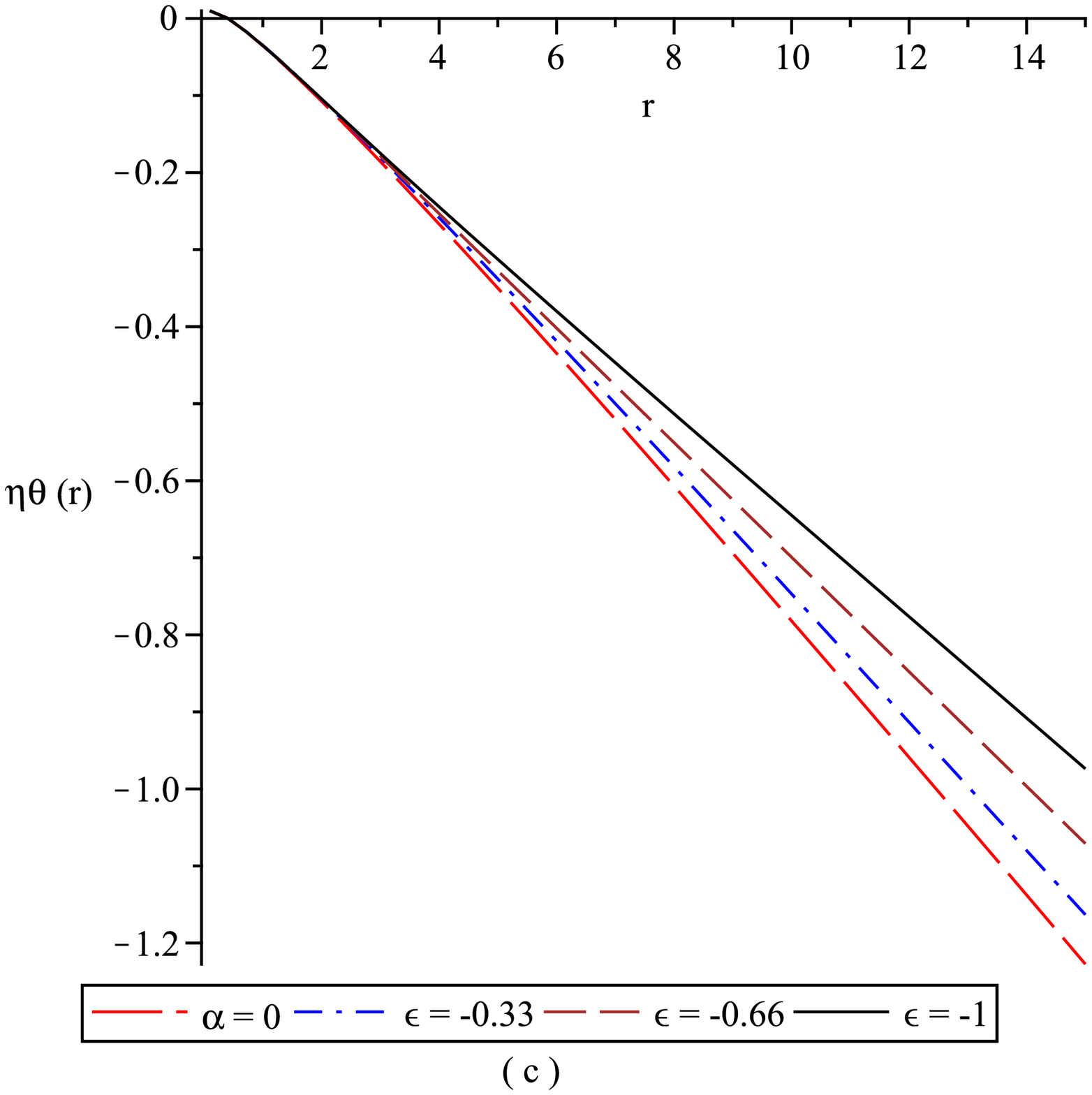}
\label{d3}
\nonumber
\end{figure}
\begin{figure}[hbtp]
\centering
\includegraphics*[width=8cm,height=6.5cm]{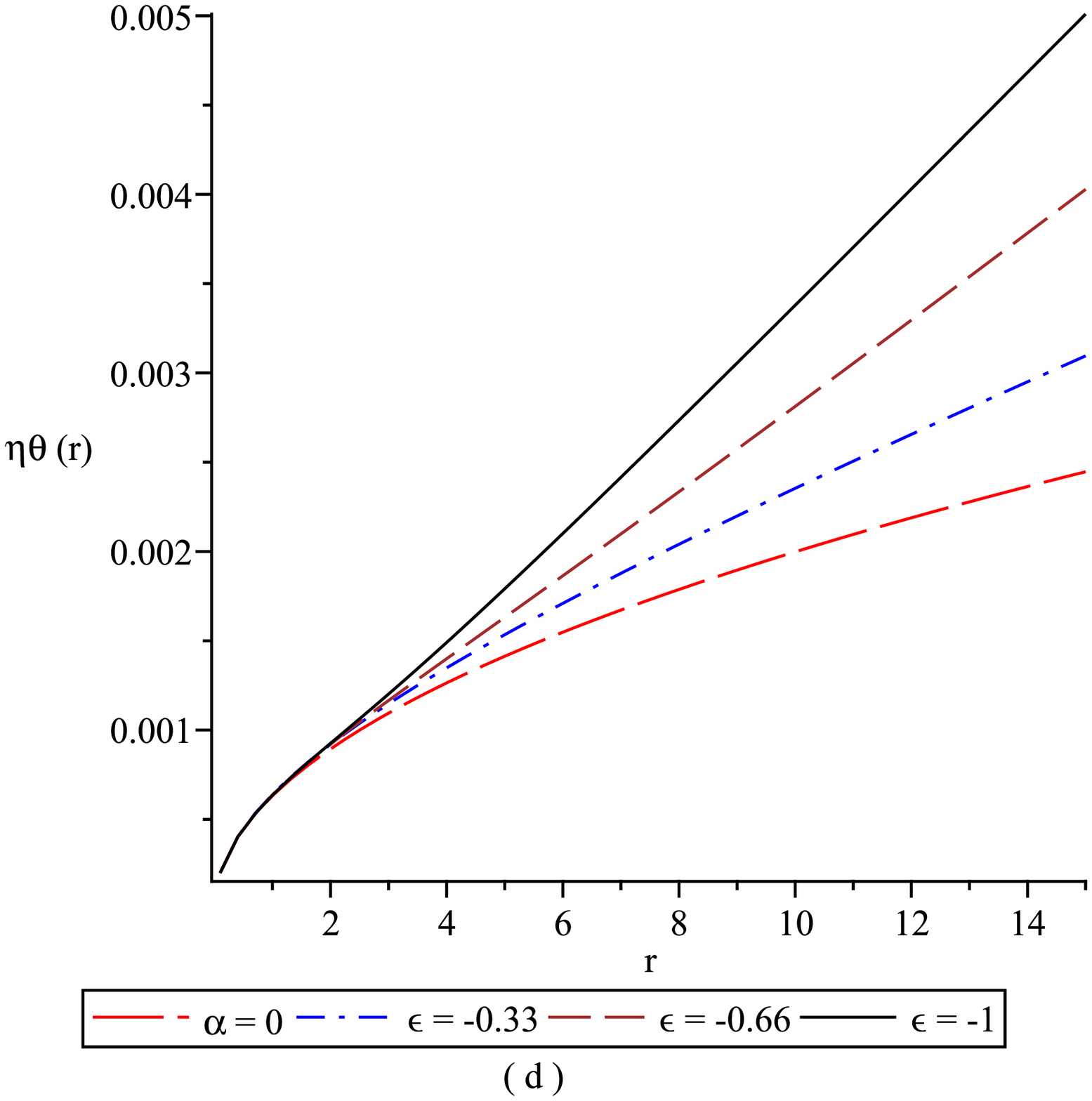}
\label{d5}
\caption{The radial variation of ${\eta}^r(r)$ and ${\eta}^{\theta}(r)$ for  different values of $\epsilon$ (label in figure) with different initial conditions:
(a) ${\eta}^r\,(0.1)=0.0002$, $\frac{d{\eta}^r\,}{dr}(0.1)=-0.001$;
(b) ${\eta}^r\,(0.1)=0.01$, $\frac{d{\eta}^r\,}{dr}(0.1)=0.001$;
(c) ${\eta}^{\theta}\,(0.1)=0.01$, $\frac{d{\eta}^{\theta}\,}{dr}(0.1)=-0.001$;
(d) ${\eta}^{\theta}\,(0.1)=0.0002$, $\frac{d{\eta}^{\theta}\,}{dr}(0.1)=0.001$. We set
 $\alpha=0.1$ and the mass of black-hole, $M=1$.}
\label{fig:deviation}
\end{figure}
In fig. (\ref{fig:deviation}), we show the behaviour of radial as well as transverse components of deviation vector $\eta(r)$, for some specific initial conditions, quoted in the same figure. One can see from fig. (\ref{fig:deviation}a) that an initially converging geodesics can diverge apart with $r$. It is also seen that with increasing the negative values of $\epsilon$ (say $\epsilon$ = -1 case),
$\eta(r)$ first reduces and then increases further. On the other hand, in case of initially diverging geodesics (see fig. (\ref{fig:deviation}b))they still remain divergent even in the presence of quintessence but at different rate. From fig. (\ref{fig:deviation}c) and fig. (\ref{fig:deviation}d), the evolution of transverse component of deviation vector $\eta(\theta)$ shows that an initially converging geodesic will converge further along radial direction but rate of convergence will reduce as $\epsilon$ increases. While an initially diverging geodesic will get diverge further with $r$. However, the effect of quintessence parameter in the transverse component of deviation vector is showing opposite as compared to the case of radial component.
Hence, in the transverse direction, as the negative value of $\epsilon$ increases, the rate of divergence also increases along with radial coordinate $r$.
\section{\label{sec:level2}Summary, Conclusions and Future Directions} \label{conclu}
In this article, we have investigated the geodesic motion for time-like geodesics in
the background of Schwarzschild black hole surrounded by quintessence field. Some of the important results are summarised below:
\begin{itemize}
\item[(i)] The effective potential for Schwarzschild black hole with quintessence contains an extra term which depends on the behavior of scalar field.
Therefore, the nature of effective potential and thus the geodesic trajectories depend on the black hole parameters as well as quintessence parameter. For a large negative value of quintessence parameter $\epsilon$, the potential does not have minima which clearly indicates the absence of a stable circular orbit for test particle in such case.

\item[(ii)] In case of non-radial geodesics, there exists all possible motion of orbits, i.e circular bound orbits (stable and unstable), radially plunge and fly-by types of orbits whereas no bound orbits are observed for a test particle in case of time-like radial geodesics. The effect of quintessence field is more enhanced for relatively smaller angular momentum ($L$) in all the cases.

\item[(iii)] As an artifact of the presence of quintessence field, the stable circular orbit of the test particle has a larger radius compare to that of Schwarzschild black hole without quintessence. Similarly, in case of unbound orbits the test particles from larger distance are set into circular motions around center, as negative value of $\epsilon$ increases.

\item[(iv)] The innermost stable circular orbits (ISCOs) are also set at larger radii as the value of $\epsilon$ becomes more negative. 
Thus it is required to restrict parameter $\alpha$ to obtain a real value of ISCO radius for a particular value of $M$ and $\epsilon$.

\item[(v)] In radial direction, quintessence assists the divergence of geodesics.
Even initially converging geodesics may diverge further if negative value of $\epsilon$ increases. While this attribute is absent in transverse direction. However, initially diverging geodesics diverge further in both the radial and transverse directions, but rate of divergence increases with the increase of negativity in the $\epsilon$ value for transverse direction.

\item[(vi)] At the singularity $r=0$, the radial component ${\eta}^r$ of geodesic deviation vector
becomes infinite, while the transverse components ${\eta}^{\theta}$ and ${\eta}^{\phi}$ vanish.
We observe that the behaviour of geodesic deviation vector in the space-time used is qualitatively similar to Schwarzschild black hole but quantitatively it differs in both the radial as well as transverse directions as presented in the representative plots (see fig. (\ref{fig:deviation})).
\end{itemize}
In view of the dynamical nature of quintessence field, it would be meaningful to perform this exercise by solving the dynamical equation of motion of scalar field to have more useful insights.
It would also be interesting to study the kinematics of geodesic flows in the background of Schwarzschild black hole with quintessence.
 For this, one needs to solve the Raychaudhuri equations for corresponding expansion, shear and rotation (or ESR) variables as an {\it initial value problem}. 
 Further, the study of the formation of accretion disks around such black holes would be important astrophysically in view of the permissible range of parameters $\epsilon$ and $\alpha$ for ISCOs.
 We hope to report on these issues in future.

\begin{acknowledgements}
The authors are indebted to the anonymous referees for useful suggestions and comments
on the manuscript which helped us to improve the presentation of the paper
significantly.
One of the authors HN would like to thank Department of Science and Technology,
New Delhi for financial support through grant no. SR/FTP/PS-31/2009.  NCD acknowledges Conselho Nacional de Desenvolvimento Cient\'ifico e Tecnol\'ogico (CNPq) for financial support provided during this work.
HN is also thankful to IUCAA, Pune for support under its visiting associateship program and RU also acknowledges the  support from CTP,  JMI, New Delhi under its visitors program during the course of this work.
\end{acknowledgements}




\end{document}